\documentclass[aps,physrev,twocolumn,superscriptaddress]{revtex4-2}

\usepackage{amsfonts}
\usepackage{amsmath}
\usepackage{booktabs}
\usepackage{graphicx}
\usepackage{makecell}
\usepackage{longtable}
\usepackage{multirow}
\usepackage{siunitx}
\usepackage{xcolor}
\usepackage{tikz}
\usetikzlibrary{positioning, shapes.geometric, fit}
\usepackage[hidelinks]{hyperref}

\begin{document}

\title{Reinforcement learning for ion shuttling on trapped-ion quantum computers}

\author{Maximilian Schier}
\email[Equal contribution, order decided randomly. Contact authors: ]{schier@tnt.uni-hannover.de, lea.richtmann@aei.uni-hannover.de}
\affiliation{Institute for Information Processing (tnt), L3S, Leibniz University Hannover, Germany}
\author{Lea Richtmann}
\email[Equal contribution, order decided randomly. Contact authors: ]{schier@tnt.uni-hannover.de, lea.richtmann@aei.uni-hannover.de}
\affiliation{Institute for Gravitational Physics, Leibniz University Hannover, Germany}
\author{Christian Staufenbiel}
\affiliation{QUDORA Technologies GmbH}
\author{Tobias Schmale}
\affiliation{QUDORA Technologies GmbH}
\affiliation{Institute for Theoretical Physics, Leibniz University Hannover, Germany}
\author{Daniel Borcherding}
\affiliation{QUDORA Technologies GmbH}
\author{Michèle Heurs}
\affiliation{Institute for Gravitational Physics, Leibniz University Hannover, Germany}
\affiliation{Deutsches Zentrum für Astrophysik (DZA)}
\affiliation{Deutsches Elektronen-Synchrotron (DESY)}
\author{Bodo Rosenhahn}
\affiliation{Institute for Information Processing (tnt), L3S, Leibniz University Hannover, Germany}

\date{\today}

\begin{abstract}
Scalable trapped-ion quantum computing is commonly realized with modular chips that feature distinct zones with specific functionalities, such as storage, state preparation, and gate execution. To execute a quantum circuit, the ions must be transported between these zones. This process is called ion shuttling. To achieve reliable computation results, the shuttling process must be optimized. However, as the number of ions increases, this becomes a high-dimensional optimization problem where optimal solutions cannot be computed efficiently. We demonstrate, to the best of our knowledge, the first use of reinforcement learning (RL) for the optimization of ion shuttling. RL is well-suited for such scenarios, as it enables learning a strategy through direct interaction with the problem. We show that our RL approach outperforms current state-of-the-art heuristic techniques, yielding a reduction in shuttling operations of up to \SI{36.3}{\percent}. Furthermore, we show that our method is easily applicable to various chip architectures. Our approach offers a versatile method to study shuttling efficiency during chip design and, therefore, a highly relevant tool for future, more complex architectures.
\end{abstract}

\maketitle

\section{\label{sec:intro}Introduction}

Trapped ions are a leading platform for fault-tolerant quantum computing, combining long coherence times with high-fidelity gates~\cite{bruzewiczTrappedionQuantumComputing2019}. Qubits are encoded in long-lived internal states and manipulated with laser or microwave fields~\cite{cirac1995quantum, sorensenQuantumComputationIons1999, zarantonelloRobustResourceEfficientMicrowave2019}.

To achieve all-to-all connectivity and enable scaling to large ion numbers, quantum charge-coupled device (QCCD) architectures were proposed~\cite{kielpinskiArchitectureLargescaleIontrap2002, pinoDemonstrationTrappedionQuantum2021}. These architectures provide distinct zones, specialized for specific operations, such as state preparation, storage, gate execution, and measurement. In addition to these specialized zones, QCCDs provide a mechanism for transporting ions between zones, a process called ion shuttling.

Despite relatively long coherence times, overall performance depends on circuit execution speed. Because shuttling is often the dominant latency, minimizing ion movements is critical~\cite{mosesRaceTrackTrappedIonQuantum2023, durandauAutomatedGenerationShuttling2023, ransfordHelios98qubitTrappedion2025}. The complexity of finding an optimal ion-shuttling schedule increases exponentially with the number of ions and circuit length. Furthermore, for every distinct circuit, an optimal solution must be computed individually. Consequently, exact optimization is impractical at scale, and current compilers rely on heuristics~\cite{schmaleBackendCompilerPhases2022, sakiMuzzleShuttleEfficient2022, muraliArchitectingNoisyIntermediateScale2020, mosesRaceTrackTrappedIonQuantum2023, wuMUSSTIMultilevelShuttle2025, schoenbergerShuttlingScalableTrappedIon2025, durandauAutomatedGenerationShuttling2023, daiAdvancedShuttleStrategies2024}.

While heuristics scale, they are often suboptimal and not generally transferable to hardware changes. This motivates learning-based approaches that discover improved schedules and adapt across devices. High-dimensional, sequential decision problems without labeled targets are well-suited to reinforcement learning (RL), making RL a natural fit for shuttling optimization.

Our study targets architectures developed within Quantum Valley Lower Saxony (QVLS)~\cite{qvls}.
We train an RL agent on arbitrary interaction circuits to minimize shuttling length, yielding a general-purpose scheduler applicable to any user input. 
Evaluated on algorithms from the MQT bench repository~\cite{mqt} and on quantum volume (QV) circuits~\cite{crossValidatingQuantumComputers2019}, the agent achieves up to \SI{36.3}{\percent} fewer movements than a prior heuristic compiler~\cite{schmaleBackendCompilerPhases2022}, for problems with up to 50 ions, which is the current QVLS target capacity~\cite{qvls-q1}. 

For small instances, we also report optimal solutions from a Boolean satisfiability (SAT) solver~\cite{schoenbergerUsingBooleanSatisfiability2024}. Our RL solutions are near-optimal in most cases, and the remaining gap shrinks when allowing longer inference time. While the SAT solver requires compilation times that are not feasible for real applications, our RL agent finds solutions quickly.

To demonstrate adaptability to chip geometry, we also train our RL agent on different architectures, enabling direct comparisons of design choices and their impact on shuttling efficiency, thereby mimicking a shuttling-aware co-design process.

These results show that RL can substantially improve ion shuttling for trapped-ion systems. It outperforms state-of-the-art heuristics, attains near-optimal performance where exact evaluation is feasible, generalizes across architectures, and supports shuttling-aware hardware design, reducing the reliance on hand-crafted heuristics as systems scale.

\section{Overview of the ion shuttling problem}

\subsection{The quantum charged-coupled device (QCCD)}
In trapped-ion quantum computing, qubits are encoded via the ion's different energy levels. To manipulate the ions, they are placed in traps where electromagnetic fields confine them. One approach is to line the ions up in a linear chain. However, linear chains are not scalable because controlling and cooling the vibrational modes in long chains becomes difficult~\cite{bruzewiczTrappedionQuantumComputing2019, muraliArchitectingNoisyIntermediateScale2020}. These shortcomings led to the design of modular systems connecting multiple traps. Instead of being linear, two-dimensional devices are built in which ions are transported between different trap zones. Often, different zones are designated for specific operations. Typical zones include storage, compute (where quantum gates are executed), and readout. This approach is known as the quantum charge-coupled device (QCCD) architecture~\cite{kielpinskiArchitectureLargescaleIontrap2002}. An important difference is that the ions now have to be transported between the zones depending on the actions to be performed on them. The actual ion shuttling on the chip is performed by varying the voltages applied to the DC electrodes, which moves the location of the axial trapping potential minimum, thereby transporting the trapped ion in the desired direction~\cite{blakestadTransportTrappedIonQubits, huculTransportAtomicIons2008}.

Executing a quantum circuit on a QCCD involves several steps. First, the circuit is translated to native gates and optimized to reduce the gate count, a process known as transpilation~\cite{schmaleBackendCompilerPhases2022}. This yields a directed acyclic graph (DAG) of native operations consisting only of single- and two-qubit gates. 
The DAG encodes execution dependencies. To execute a native gate, the ions involved must be brought to the compute zone, so we require a protocol that shuttles ions through the chip’s zones while preserving the circuit’s logical order.
Gates can be reordered according to their commutation relations.
We only consider two-qubit gates as commuting if they act on distinct qubits, as in the reference methods. While this assumption is unnecessarily conservative in special cases, it allows to simplify the encoding and simulation.
Single-qubit gates can be executed immediately before the two-qubit operation, once the ions are already in the compute zone. Accordingly, we temporarily remove all single-qubit gates before optimizing shuttling. 

After a shuttling protocol is synthesized, the single-qubit gates are reinserted into the schedule. Throughout this work, we use “qubit” and “ion” interchangeably.

\subsubsection{Example architecture 1: X-chip}
Our first example architecture is the QVLS QROSS chip~\cite{ungerechts_designing_}, the first proposal for a QCCD developed by Quantum Valley Lower Saxony (QVLS)~\cite{qvls}. It consists of four registers connected by an X-junction, we refer to it as the ``X-chip''. The registers include a compute zone that can hold up to 2 ions, a state preparation and measurement (SPAM) zone that can hold only 1 ion, and two storage regions. An abstract sketch of this design is shown in the upper left corner of Figure~\ref{fig:representation}.  The registers exhibit a stack-like behavior: when the ion closest to the junction is extracted from a zone and moved through the junction to another zone, the ion behind it is simultaneously moved next to the junction. This means that each time an ion crosses the junction, it counts as one shuttling step.  

The maximum number of ions a storage register can hold determines the maximum number of qubits available for computation. Simulations for the X-chip were run with up to 50 ions; the first fabricated chip will support up to 24 ions~\cite{ungerechts_designing_}.

\begin{figure}[h]
    \centering
    \includegraphics[width=0.5\linewidth]{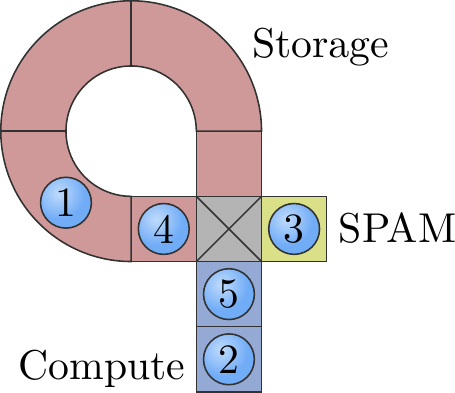}
    \caption{\label{fig:q-chip}Layout of the CIRQLE ion trap chip, also called Q-chip. The shown sketch has a single storage ring with a capacity of 5 qubits, which is connected to the other two registers with a junction. This ring acts like a carousel with empty or filled spots: it can only transport all stored qubits simultaneously through rotations. Some operations may only move empty spots through the junction. For example, in the given configuration, a clockwise rotation of the storage element does not move any qubit through the junction. It is likely that these operations are faster on real hardware than shuttling qubits through the junction. Our proposed method can handle this due to its Semi-Markovian formulation, unlike other methods such as the SAT solver.}
\end{figure}

\subsubsection{\label{sec:Q-chip}Example architecture 2: Q-chip}

We also study an alternative chip design, the QVLS CIRQLE chip~\cite{ungerechts_designing_2026}, with a more compact storage register, allowing more ions to fit on the same chip size. 
Here, the storage zone is consolidated in a ring. The compute zone (capacity 2 ions) and the SPAM zone (capacity 1 ion) are connected to this ring via a junction, resulting in a Q-like shape, which is why we refer to it as the ``Q-chip'', see Figure~\ref{fig:q-chip}. While in the X-chip, ions can be moved in and out of one storage zone independently of the other storage zone, in the Q-chip, we have only one common storage zone, and each movement will affect the location of all the other ions in the ring. 
The Q-chip storage ring works like a carousel with spots, some occupied, some empty. It can only rotate as a whole in either direction. To move an ion to a different zone, the ring is rotated until that ion is next to the junction and then transported to the destination zone.

Rotating about a single spot, whether it is filled or not, counts as one shuttling action. However, rotating empty spots through the junction is considered less costly than rotating ions through the junction, as there is no risk of ion loss. This non-uniform step duration is accounted for in our algorithm later. 

\subsection{\label{sec:MotivatoinOptimizing}Importance of optimizing the shuttling operations}

The execution time of a quantum circuit is dominated by ion shuttling, which often accounts for more than half of the total runtime~\cite{mosesRaceTrackTrappedIonQuantum2023, durandauAutomatedGenerationShuttling2023, ransfordHelios98qubitTrappedion2025}. 

The main reason for reducing shuttling times is decoherence. While trapped ions offer relatively long coherence times - in the order of minutes for hyperfine qubits~\cite{bruzewiczTrappedionQuantumComputing2019} - current transport operations still take milliseconds or microseconds~\cite{muraliArchitectingNoisyIntermediateScale2020, pinoDemonstrationTrappedionQuantum2021, ransfordHelios98qubitTrappedion2025}. The execution of large quantum circuits with high qubit counts can quickly exceed 10,000 shuttling operations on current QCCD architectures and can therefore take longer than the qubits' coherence times. Hence, to execute such circuits with high fidelity, optimized shuttling protocols are crucial. 

To lower transportation overhead, several methods have been studied to physically reduce time requirements of individual transport operations~\cite{bowlerCoherentDiabaticIon2012, waltherControllingFastTransport2012, luFastShuttlingParticle2018, kaushalShuttlingbasedTrappedionQuantum2020, mosesRaceTrackTrappedIonQuantum2023}. However, increasing transport speed can, in the worst case, lead to ion loss. Besides this risk, the faster the transport, the greater the ions' energy, which in turn leads to heating. This makes it more difficult to address the ions for gate execution and is detrimental to the gate fidelity~\cite{muraliArchitectingNoisyIntermediateScale2020, kaushalShuttlingbasedTrappedionQuantum2020, sakiMuzzleShuttleEfficient2022, durandauAutomatedGenerationShuttling2023, wuMUSSTIMultilevelShuttle2025}. To remove the energy acquired, an extra cooling step can be implemented~\cite{ransfordHelios98qubitTrappedion2025}. Still, for achieving high gate fidelities, the transport speeds are limited, and the amount of transport steps should therefore be reduced as much as possible~\cite{kielpinskiArchitectureLargescaleIontrap2002, sakiMuzzleShuttleEfficient2022}. 

Additional decoherence can occur during transport due to inhomogeneous magnetic fields~\cite{bruzewiczTrappedionQuantumComputing2019}. While modern shuttling techniques mitigate these effects, they remain a potential source of error; optimizing the shuttling schedule reduces the qubits’ exposure time~\cite{mosesRaceTrackTrappedIonQuantum2023, ransfordHelios98qubitTrappedion2025}.

Overall, reducing the shuttling times is widely considered crucial for achieving reliable trapped-ion quantum computing with QCCDs~\cite{bruzewiczTrappedionQuantumComputing2019, ransfordHelios98qubitTrappedion2025}. 

\subsection{\label{sec:CurrentTechniques}State-of-the-art shuttling techniques}

As the amount of available qubits in current QCCDs increases, it is usually not possible to develop scalable methods that are able to find an optimal solution to minimize the shuttling operations~\cite{sakiMuzzleShuttleEfficient2022}. Methods for finding shuttling protocols, therefore, often rely on sorting techniques~\cite{pinoDemonstrationTrappedionQuantum2021, mosesRaceTrackTrappedIonQuantum2023} and on heuristics to determine which ions are shuttled in what order. Current strategies for ion shuttling differ widely in terms of the types of architectures they study (amount of storage and gate zone, connections between these zones, possibility to perform parallel operations) and shuttle operations they include (linear shifts, movements through junctions, splitting and combining ion crystals, and swapping the positions of two ions). As a summary, we give here an overview of the main types of heuristics that are applied:\\
\textit{Determination of possible tasks:} This refers to the heuristic where, instead of looking at all the tasks, i.e., gates to be executed, the compiler focuses on the tasks that can actually be performed, making the action space smaller~\cite{muraliArchitectingNoisyIntermediateScale2020}. Approaches often use a dependency graph in order to determine the order in which gates have to be executed~\cite{sakiMuzzleShuttleEfficient2022, mosesRaceTrackTrappedIonQuantum2023, wuMUSSTIMultilevelShuttle2025} and to select which gate should be executed next~\cite{schmaleBackendCompilerPhases2022, schoenbergerShuttlingScalableTrappedIon2025}.\\
\textit{Least movements:} This group of heuristics prioritizes gates that can be executed quickly, when one ion is already in the compute zone or moves ions first that are close to the compute zone~\cite{schmaleBackendCompilerPhases2022, mosesRaceTrackTrappedIonQuantum2023, wuMUSSTIMultilevelShuttle2025}.\\
\textit{Storage selection:} This heuristic refers to making selective use of different storage regions:  \cite{wuMUSSTIMultilevelShuttle2025} move qubits that are not needed in the near future to a designated storage location so they do not block the path. In a similar idea, \cite{mosesRaceTrackTrappedIonQuantum2023} store ions that are used together in proximity. \cite{schmaleBackendCompilerPhases2022} also implement proximity sorting and use the readout zone as temporary storage to make ions needed shortly after easily accessible.\\
\textit{Handling traffic blocks:} Different heuristics are used in order to avoid an ion from being blocked because the path it has to take is occupied by other ions~\cite{wuMUSSTIMultilevelShuttle2025}. This is done, for example, through specific movement combinations like splitting and merging~\cite{muraliArchitectingNoisyIntermediateScale2020, sakiMuzzleShuttleEfficient2022} or by moving ion chains on closed paths together in loops~\cite{schoenbergerShuttlingScalableTrappedIon2025, schoenbergerOrchestratingMultiZoneShuttling2025, schoenbergerShuttlingTrappedIonQuantum2025} for architectures where this is possible.

\cite{daiAdvancedShuttleStrategies2024} propose a probabilistic formula that accounts for several heuristics, such as \textit{least movements} and \textit{handling traffic blocks}. With this formula, they can then determine whether an ion should move or stay in the same trap. 

The mentioned strategies are planning strategies; a shuttling protocol is developed for a given circuit in advance and then executed on the actual QCCD. \cite{ransfordHelios98qubitTrappedion2025} in contrast use real-time compilation. They also employ heuristics, such as allocating storage regions and accounting for movement costs.

For the sake of concreteness, we compare our proposed RL-based approach against two implementations of the above-mentioned methods:
the heuristics-based compiler based on~\cite{schmaleBackendCompilerPhases2022}, and the SAT-based approach from~\cite{schoenbergerUsingBooleanSatisfiability2024, schmale2026hybrid}.
We chose these two approaches for our comparisons because they have been used so far for the QVLS X-chip.

\subsubsection{Reference  1: Heuristic compiler}
In the framework of the QVLS X-chip, a shuttling compiler was developed to address the ion shuttling problem using heuristics derived from observations of the chip’s architecture~\cite{schmaleBackendCompilerPhases2022}. The challenge of orchestrating ions across the chip to execute a given quantum circuit was therefore decomposed into several phases, two of which are particularly relevant to this paper:

\textbf{Graph serialization:} The compiler begins by representing the input quantum circuit as a directed acyclic graph (DAG), which encodes a partial order for gate execution. This partial order allows for flexibility in scheduling, enabling the compiler to select an execution sequence that minimizes the number of ion shuttling operations. A custom algorithm generates $n_p$ distinct sequences of up to $n_g$ interaction gates (i.e., two-qubit gates), each respecting the partial order. The sequence requiring the fewest shuttling operations (determined via the subsequent compiler stages described below) is selected for execution, removed from the DAG, and executed. This process iterates until the DAG is empty.

\textbf{Ion orchestration:} Given a selected sequence of $n_g$ interaction gates, the ion orchestration module generates a sequence of shuttling directives to move ions across the chip. The primary objective is to repeatedly position ions into the compute zone such that the $n_g$ gates can be executed in the prescribed order. Different heuristics are employed to guide the generation of efficient shuttling sequences. For details on the heuristics, see~\cite{schmaleBackendCompilerPhases2022}.

Increasing the size of the intermediate gate sets ($n_g$) allows the compiler to consider more future gates, thereby reducing the total number of shuttling operations. However, this comes at the cost of increased compilation time. For this work, we set $n_g=4$, which provides a favorable trade-off between short shuttling sequences and reasonable compile time.

This work employs a version of the heuristic based compiler developed further by QUDORA Technologies GmbH. While minor enhancements have been made to the original release, the core structure and heuristics of the shuttling compilation process remain unchanged compared to the original publication~\cite{schmaleBackendCompilerPhases2022}.

\subsubsection{Reference 2: SAT solver}
For benchmarking purposes, it is useful to compare obtained trajectories against optimal ones. While finding these optimal trajectories is likely unfeasible in the general case, we can at least study small instances to gain some basic insight. 

In principle, a naive exhaustive search through all shuttling sequences of a fixed length could be used to find the shortest option for a given input circuit, however, a simple estimate shows that this becomes infeasible quite quickly: For some of the QCCD architectures studied below, there are at least 3 possible next shuttling operations from any given ion configuration. For some of the below circuits, we find optimal shuttling sequences consisting of more than 70 steps, yielding a total of more than $3^{70}\approx10^{33}$ shuttling sequences which the naive search would have to consider.
\cite{schoenbergerUsingBooleanSatisfiability2024} suggests to ameliorate this situation by formulating the ion shuttling problem as a boolean satisfiability problem (SAT) and leveraging the decades-long efforts that have gone into the development of automated solvers for such problems (``SAT solvers'').

The core idea is to define boolean variables $b_{i, x, t}$ that indicate whether an ion $i\in\{0, ...N\}$ is located at location $x$ at time step $t\in\{0, ..., T\}$, for some upper limit on the number of time steps $T$ and ion count $N$. For a given circuit and chip geometry, one can then construct a boolean function $\Phi_T: \{b_{i, x, t}\} \rightarrow \{\text{True}, \text{False}\}$ that encodes whether the input variables $\{b_{i, x, t}\}$ encode a valid shuttling sequence for executing the circuit. For details on how $\Phi_T$ is constructed, see~\cite{schoenbergerUsingBooleanSatisfiability2024, schmale2026hybrid}. Finding such a set of variables that makes $\Phi_T$ evaluate to ``True'' (``satisfying assignment'') is precisely an instance of a SAT problem.
In such a representation, existing solvers can be used to check whether a satisfying assignment exists for a particular value of $T$. If the solver indicates that $\Phi_T$ is not satisfiable (i.e. no valid shuttling sequence of length $T$ exists), $T$ may be incremented until a solution is found. If this procedure is started from $T=1$, a provably optimal solution will be found. From a satisfying assignment, a shuttling schedule can then be reconstructed.

This approach is not scalable, since SAT is NP-complete in general~\cite{sipser1996introduction}. However, we find this approach to work very well in practice for small problems, allowing us to optimally compile instances unapproachable using naive exhaustive search.

\subsection{Motivation for using reinforcement learning for ion shuttling}

As explained in Section~\ref{sec:MotivatoinOptimizing}, it is paramount for reliable trapped-ion quantum computing to reduce the ion shuttling overhead. For large numbers of ions, this becomes a high-dimensional optimization problem, making it impractical to find optimal solutions. This is why, as shown in Section~\ref{sec:CurrentTechniques}, heuristics are employed.
However, by their nature, heuristic approaches cannot comprehensively cover every eventuality, and efforts to achieve such completeness lead to a disproportionate increase in development effort.
High-dimensional optimization problems that are intractable for conventional optimization methods are a classical use case for machine learning. 
Because the optimal solution is unknown, labeled data are unavailable, and supervised learning is not applicable. Reinforcement learning, on the other hand, is well-suited: an agent takes sequential actions, observes the resulting state, and receives a reward reflecting progress toward the objective. Initially, the agent explores randomly. It updates its policy, that is, its rules for decision making, to maximize its cumulative reward and progressively learns efficient strategies.
It is therefore also capable of finding patterns that were not thought of theoretically beforehand. By training the RL agent on a diverse set of randomly generated circuits, it learns to generalize and compile arbitrary circuits without retraining.

Reinforcement learning has been used previously to optimize the transpilation process~\cite{bukovReinforcementLearningQuantum2026}. However, to the best of our knowledge, this is the first work to employ it for the optimization of ion shuttling.

\section{\label{sec:proposed}Proposed Reinforcement Learning Method}
In this section we present the relevant concepts of reinforcement learning as well as the important technical details of our proposed RL ion shuttling compiler, which we call ``RLIonS''.

\subsection{Problem formulation}
We formulate ion shuttling as a Semi-Markov Decision Process (SMDP)~\cite{puterman2014markov}. Like a regular Markov Decision Process (MDP)~\cite{SuttonBarto2018RL}, the SMDP models the interaction of an agent with a system at certain time points, the discrete ``events''. Unlike an MDP, an SMDP permits the system to behave non-Markovian while transitioning between time points under continuous time. More importantly, the SMDP formulation allows that steps between two events have non-uniform duration, the standard MDP does not. This is an important aspect when modeling an arbitrary ion shuttling chip, as some shuttling operations may take longer than others.

Formally, an SMDP is defined as the tuple $(\mathcal{S}, \mathcal{A}, A, P, F, R, \beta)$. 
\begin{itemize}
    \item The state space $\mathcal{S}$ is the discrete space of all possible configurations of the ion shuttling simulation. This entails the location of all qubits on the trapped-ion chip and the pending two-qubit gates.
    \item The action space $\mathcal{A}$ is the discrete space of the ion movements the chip can perform in general.
    \item The action mask $A : \mathcal{S} \rightarrow \mathcal{P}(\mathcal{A}) \setminus \{\emptyset \}$ describes the permissible subset of actions for a given state. Here, $\mathcal{P}$ is the power set.
    \item The transition dynamics $P : \mathcal{S} \times \mathcal{A} \rightarrow \mathcal{S}$ describe the new state at the next decision epoch after performing an action from a given state. Transitions are deterministic for ion shuttling.
    \item The decision epoch duration $F : \mathcal{S} \times \mathcal{A} \rightarrow (0, \infty)$ describes the duration until the next decision epoch, i.e. the duration until the agent selects the next action. It is dependent on the current state and action and is deterministic for our problem.
    \item The reward function $R:\mathcal{S} \times \mathcal{A} \times \mathcal{S} \rightarrow \mathbb{R}$ describes the immediate reward for a transition, i.e. a feedback how beneficial the selected action is during the current decision epoch. 
    \item The discount rate $\beta \in (0, \infty)$ describes how future rewards are valued. It is linked to the discount factor of a standard MDP through $\gamma = e^{-\beta}$ assuming a decision epoch duration of 1.
\end{itemize}
A (probabilistic) policy is any function of the form $\pi : \mathcal{S} \rightarrow \Delta(\mathcal{A})$, where $\Delta$ is a probability simplex. A trajectory $\zeta$ is a sequence of states and actions that can be obtained by rolling out the interaction of a policy and the system described by the SMDP: $\zeta = (s_0, a_0, s_1, a_1, \ldots)$. The initial state $s_0$ is usually sampled from a distribution of valid start states $s_0 \sim \rho$, with $\rho : \Delta(\mathcal{S})$. Performing a rollout using a policy $\pi$ is commonly notated in short form $\zeta \sim \pi$. The total discounted return $G$ of a trajectory, i.e., the discounted sum of the individual step rewards, is defined as:
\begin{equation}
G(\zeta) = \sum_t e^{-\beta \cdot \sum_{u=0}^{t-1} F(s_u, a_u)} R(s_t, a_t, s_{t+1})\,\text{.}
\end{equation}
Given such an SMDP, the goal is finding the optimal policy $\pi^*$ over the expected discounted return $J$:
\begin{align}
\pi^* &= \operatorname{arg}\operatorname{max}_{\pi} J(\pi)\,\text{, where} \\
J(\pi) &= \mathbb{E}_{\zeta \sim \pi} [G(\zeta)]\, .
\end{align}

Since the goal in ion shuttling is finding the shortest valid shuttling sequence, the reward term should continuously penalize at a constant rate $c_r$ for the entirety of a decision epoch:
\begin{align}
    R(s, a, s') &= \int_0^{F(s, a)}-c_r\cdot e^{-\beta t}\operatorname{d}t = c_r\cdot\left[\frac{e^{-\beta t}}{\beta}\right]_0^{F(s, a)} \\ &= c_r\cdot\frac{e^{-\beta F(s, a)} - 1}{\beta} \,\text{.}
\end{align}

\subsection{Proximal Policy Optimization}
A parameterized policy is defined as $\pi_\theta : \mathcal{S} \rightarrow \Delta(\mathcal{A})$, with parameters $\theta$. The policy gradient theorem~\cite{sutton1999policy} specifies that $\pi_\theta$ converges locally by gradient ascent on:
\begin{equation}
    \nabla_{\theta} J(\pi_{\theta}) = \mathbb{E}_{s \sim d^{\pi_{\theta}}, a \sim \pi_{\theta}(\cdot | s)}\left[\Psi^{\pi_{\theta}}(s, a) \nabla_{\theta} \operatorname{log} \pi_{\theta}(a|s)\right]\,
\end{equation}
if $\Psi^{\pi_\theta}$ is a bias-free estimator of the state-action values or advantages of $\pi_{\theta}$. Here, $d^{\pi_{\theta}}$ is the discounted state distribution under policy $\pi_\theta$. This objective requires that all sampled trajectories originate from the current policy $\pi_\theta$. Thus, after every gradient step, an often impractically large number of trajectories must be collected for the next step.

Proximal Policy Optimization~\cite{schulman2017proximal} is a commonly used policy gradient method that reduces the number of samples per gradient step. First, it is allowed that samples used for training are collected by a close policy but not exactly the same policy. This is corrected using importance sampling and clipping. Let $\theta'$ be an old close parameterization of $\pi_\theta$. The objective of PPO is defined as:
\begin{equation}
    \begin{split}
        \nabla_{\theta}J(\pi_{\theta}) = \nabla_{\theta} \mathbb{E}_{s \sim d^{\pi_{\theta'}}, a \sim \pi_{\theta'}(\cdot|s)} \biggl[\operatorname{min} \biggl(\Psi^{\pi_{\theta'}} \frac{\pi_{\theta}(a|s)}{\pi_{\theta'}(a|s)}, \\ \Psi^{\pi_{\theta'}} \operatorname{clip} \biggl( \frac{\pi_{\theta}(a|s)}{\pi_{\theta'}(a|s)}, 1 - \epsilon, 1 + \epsilon \biggr)\biggr) \biggr] \,\text{,}
    \end{split}
\end{equation}
where $\epsilon$ is the clipping ratio.
Second, the estimator $\Psi^{\pi_{\theta'}}$ of PPO is biased. PPO uses a combination of the bias-free estimator given by the Monte-Carlo rewards and the biased estimator of a parameterized model $V_\phi : \mathcal{S} \rightarrow \mathbb{R}$. For a regular MDP, PPO uses the Generalized Advantage Estimator (GAE) $\hat{A}_t$~\cite{schulman2015high} as $\Psi^{\pi_{\theta}}$, which is defined as:
\begin{align}
    \hat{A}_t &= \delta_t + (\gamma\lambda)\hat{A}_{t+1}\,\text{, where}\\
    \delta_t &= R(s_t, a_t, s_{t+1}) + \gamma V_{\phi}(s_{t+1}) - V_\phi(s_t)\,\text{.}
\end{align}
For an SMDP, the GAE is modified to:
\begin{align}
    \hat{A}_t &= \delta_t + (\lambda e^{-\beta})^{F(s_t, a_t)}\hat{A}_{t+1}\,\text{, where}\\
    \delta_t &= R(s_t, a_t, s_{t+1}) + e^{-\beta F(s_t, a_t)} V_{\phi}(s_{t+1}) - V_\phi(s_t)\,\text{.}
\end{align}

\subsection{Representation}
\begin{figure*}
    \centering
    \includegraphics[width=1.0\linewidth]{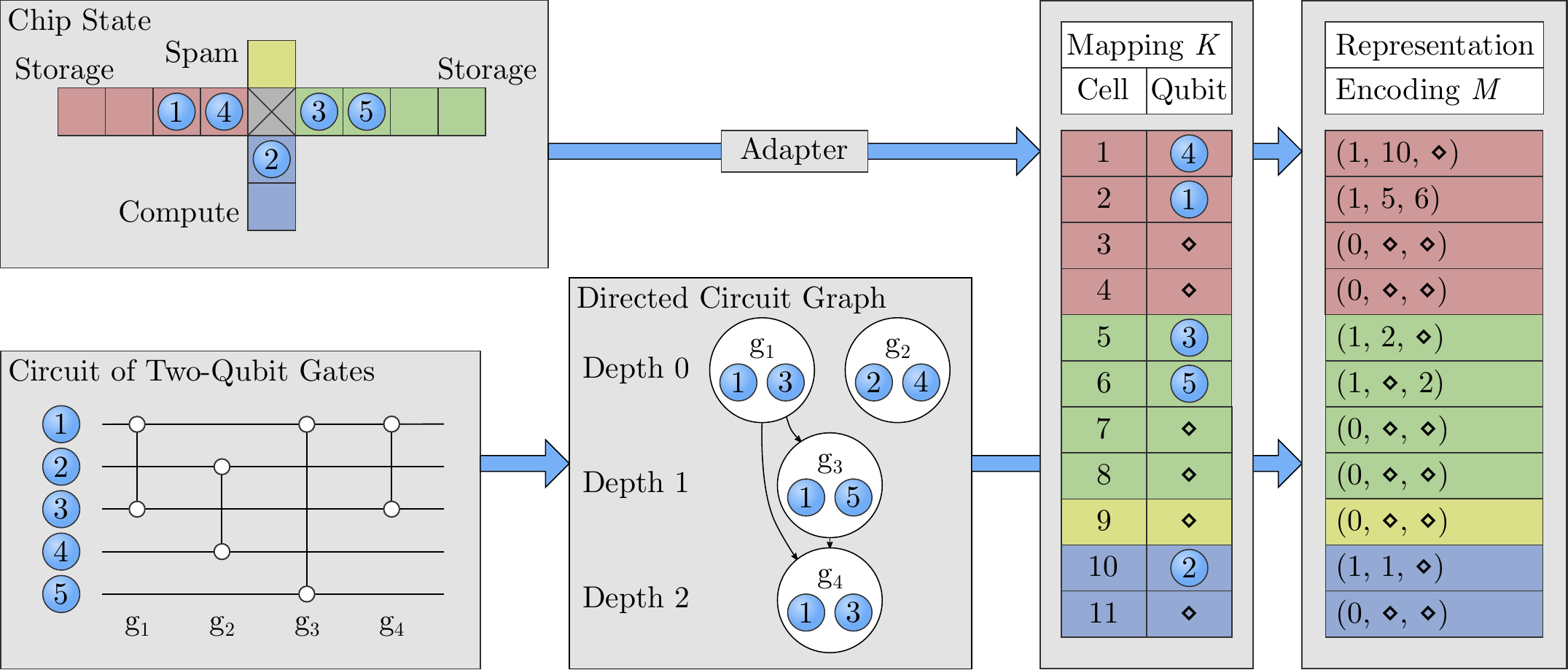}
    \caption{\label{fig:representation}Proposed representation. The top left shows the chip state. A chip-specific adapter translates the chip state into a cell-qubit mapping, shown in the mapping table $K$ on the right side. The rest of the observation encoding is chip-agnostic. 
    The directed acyclic graph of two-qubit gates (middle) is constructed from the circuit (bottom left).
    The depth of each gate is determined. Finally, the encoding table $M$ is constructed. Each row refers to a cell on the given chip. The first entry per row indicates presence of a qubit ($0$ if the cell is empty). 
    The next entries within a row indicate the cell location of the other operands at increasing depth up to the lookahead. The special value $\diamond$ indicates no gate at this depth.
    For example, consider the encoding of cell $6$. This cell holds qubit $(5)$, which is used only in gate $g_3$ located at a depth of 1. That cell's encoding is $(1, \diamond, 2)$, because it has a qubit present (1 in first entry), that qubit has no gate at depth 0 (therefore the $\diamond$ in second entry), but it does have a gate at depth 1, which uses the qubit $(1)$ as the other operand, which is located in cell $2$ shown in the third entry. Depth 2 is not considered for this example with $k_{\operatorname{lookahead}} = 2$. Note that in our training we implement a lookahead of 4.}
\end{figure*}

A trapped-ion chip can have a geometrically complex shape, but ultimately the ions acting as qubits are located at positions from a discrete set. We can therefore express the state of a given chip in tabular form. Let $n_{\text{cells}}$ be the number of positions any ion can be located at, which we call ``cell''. The configurations of any chip is then expressed by a mapping or table from cells to qubits $K: \{1, \ldots, n_{\text{cells}}\} \rightarrow \{\diamond, 1, \ldots, n_{\text{max}}\}$, where $\diamond$ indicates an empty cell and $n_{\text{max}}$ is the maximum number of ions supported by a given chip. A valid mapping should also have no duplicate qubits: $\forall i, j : i \neq j \implies K(i) = \diamond \lor K(i) \neq K(j)$.
How many cells exist and how they are mapped to the physical locations of the chip depends on the chip at hand, but is usually straightforward. An example for the QVLS X-chip is shown in Figure~\ref{fig:representation}.
Next, we assume that a circuit of two-qubit gates is given as a sequence of gates $p = (g_1, g_2, \ldots)$, with gate $g_i = (x_i, y_i)$. Thus, $x_1$ is the first operand of the first gate and $y_1$ the second operand of the first gate and so on. 
Gates do not have to be executed in sequence, as long as the gate to be executed has no operand used in a prior gate that still needs to be executed.

\subsubsection{Requirements for representations}
When employing a neural network as a policy $\pi$, the state space $\mathcal{S}$ must usually be transformed to compatible representations, e.g. vectors, using some transformation. For simplicity in a slight abuse of notation we use $\mathcal{S}$ as the representation space directly.
A good representation of the chip state and circuit to be executed should be invariant to changes which have no influence on the validity of any given trajectory and the return under that trajectory and are thus equivalent regarding the solution.
In the RL literature, such a relation of states is called a bisimulation relation~\cite{givan2003equivalence}. 
Let $B_M$ be the maximal bisimulation relation, such that the partition $\mathcal{S} / B_M$ has lowest cardinality. A good state representation $\mathcal{S}$ should minimize $\lvert \mathcal{S} \rvert - \lvert \mathcal{S} / B_M \rvert$.
Otherwise, an agent must first learn that an infeasible large number of representations all represent the same underlying state.

\paragraph{Invariance to relabeling}
First, a good representation should be invariant to a relabeling of the qubits. Such a relabeling can be carried out by swapping the label of two qubits in $K$ and also swapping all occurences in $p$.

\noindent
\emph{Example:} Consider an arbitrary chip state with 50 qubits. There are $50! \approx 3 \cdot 10^{64}$ unique relabeling permutations of the state.

\noindent
\emph{Conclusion:} A good representation should not directly contain the label of a qubit.

\paragraph{Invariance to commutative circuit reordering}
As described above, in the sequence of two-qubit gates $p$, the first gate does not need to be executed first, as long as another gate after it has two operands that do not appear before that gate.
Thus, any two neighboring gates $g_i=(x_i, y_i)$ and $g_{i+1}=(x_{i+1}, y_{i+1})$ can be swapped and the resulting new circuit is equivalent, as long as $x_i \neq x_{i+1} \land x_i \neq y_{i+1} \land y_i \neq x_{i+1} \land y_i \neq y_{i+1}$.

\noindent
\emph{Example:} Consider the QV(50) problem. There are $50$ independently generated layers of $25$ two-qubit gates each. Within a layer all two-qubit gates are commutative. Due to the problem structure, there are at least $(25!)^{50} \approx 3 \cdot 10^{1259}$ equivalent circuit permutations for any given start state.

\noindent
\emph{Conclusion:} A representation should not directly contain the sequence index $i$ of a two-qubit gate $g_i$.

\subsubsection{\label{sec:proposed_representation}Proposed representation}
The core idea of our proposed representation is abstracting the qubit label and sequence position of a two-qubit gate by encoding a gate through the cell-location of the other operand and the depth of the gate in the dependency graph. This is illustrated in Figure~\ref{fig:representation} for a lookahead of $k_{\operatorname{lookahead}} = 2$. The following steps are performed:
\begin{enumerate}
    \item A chip-specific adapter translates the chip state (top left) into a tabular form $K$ (columns ``Cell'' and ``Qubit'' on the right). In our case the adapter simply iterates all zones starting with the position next to the junction.
    \item If the circuit (bottom left) is given as a list of gates, the directed acyclic graph of the circuit is calculated first (bottom center). The depth of a gate is defined as the length of the longest path on the transpose graph starting from the node of that gate.
    \item The encoding matrix $M$ is computed (right). For each cell, it is encoded whether it is occupied by a qubit. Next, for depths in $\{0,\ldots, k_{\text{lookahead}} - 1\}$, it is checked if a gate at that depth exists with the qubit of the current cell. If it exists, the cell of the other operand is encoded. Otherwise, an empty token $\diamond$ is encoded.
\end{enumerate}
Gates at a depth of $k_{\text{lookahead}}$ or larger are not included in the representation. This is the case for $g_4$ in Figure~\ref{fig:representation}. The final observation input for the agent is constructed by flattening $M$ and appending the total remaining gate count.
We then employ a sinusoidal embedding~\cite{vaswani2017attention} to encode numeric values, see Appendix~\ref{sec:appendix_sinusoidal} for details.

\subsection{Network architecture}
Both the policy and value function are implemented as dense neural networks with residual feedforward blocks. Such a residual feedforward block of width $h$ applies in sequence a ReLU activation, a fully connected layer with $h$ units, another ReLU, and another fully connected layer with $h$ units. Finally, in the residual feedforward block, the initial input signal is added to the output again. Policy and value networks share no parameters. Per network, we use one fully connected layer to project onto the hidden feature size $n_{\operatorname{hidden}}$. Next, $n_{\operatorname{blocks}}$ residual blocks of width $h = n_{\operatorname{hidden}}$ are applied in sequence. Finally, another ReLU activation and a last fully connected layer is applied. This last layer has 1 output for the value function and $n_{\operatorname{act}}$ for the policy. This design is selected as residual connections are a common choice in current RL literature~\cite{espeholt2018impala,lee2025simba}.

\subsection{Description of training process}
\subsubsection{\label{sec:shaped_reward}Shaped reward}

The basic reward signal for a goal-reaching problem is very sparse, as the agent receives a negative reward at a constant rate $c_r$ until a goal state is reached. If the problem only terminates upon reaching a goal state and the agent has not encountered any goal states yet, the value of every state must be estimated as ${V = -c_r\int_0^\infty e^{-\beta t} \text{d}t = -\frac{c_r}{\beta}}$. The resulting optimal policy until a goal state is encountered is a uniform distribution over the allowed actions in every state. On complex problems this is a very poor exploration policy due to the high branching factor of the graph and the resulting large number of states searched while lacking any guidance. 
To give the agent some prior knowledge, a potential-based shaped reward can be introduced. For any standard MDP, a surrogate MDP can be constructed with any reward function of the form:
\begin{equation}
    R'(s, a, s') = R(s, a, s') + \gamma_s\phi(s') - \phi(s)\, ,
\end{equation}
where $R$ is the original reward, $\gamma_s = \gamma$ is the original discount factor, and $\phi$ is some potential over the state space which is $0$ for all terminating states: ${\forall s \in \mathcal{S}_{\operatorname{T}}:\phi(s) = 0}$. When these conditions are met, any policy which is near-optimal on $R'$ is also near-optimal on the original $R$~\cite{ng1999policy}.
For the SMDP, we modify the function to account for the step duration:
\begin{equation}
    R'(s, a, s') = R(s, a, s') + \gamma_s^{F(s, a)}\phi(s') - \phi(s)\, .
\end{equation}
Let $f_{\operatorname{gates}}(s)$ be the number of gates that still need to be applied in state $s$. Then, our proposed potential-based shaped reward is:
\begin{equation}
    \phi(s) = -f_{\operatorname{gates}}(s)\, .
\end{equation}
If $\gamma_s$ is selected differently from $\gamma$, then a policy which is near-optimal on the surrogate problem is not necessarily near-optimal on the original problem. We set $\gamma_s > \gamma$, making the policy greedy in improving on the potential $\phi$. 
Greedy in this context means that the agent prefers more immediate reward.
This is actually required if the effective time horizon induced by $\gamma$ is too short to solve the problem, which can be the case for large quantum circuits as we show in our experiments (see Section~\ref{sec:ablations}).

\subsubsection{Problem  generation during training}
When training the RL agent, a diverse range of starting states is desirable, such that the entire possible problem space is well covered.
A starting state is generated by first drawing the number of ions or qubits on the chip: $z \sim \operatorname{Uniform}(\{2, \ldots, n_{\operatorname{max}}\})$.  Here, $n_{\text{max}}$ is the maximum number of ions supported.
The qubits are positioned on a random storage element. Next, we sample on average $\frac{z}{n_{\operatorname{max}}}\cdot n_{\operatorname{gates}}$ gates, where $n_{\operatorname{gates}}$ is selected to reflect the number of gates expected in a large circuit.
Per gate, the first operand $x_i \sim \operatorname{Uniform}(\{1,\ldots,z\})$ is sampled from the available qubits, then the second without replacement: $y_i \sim \operatorname{Uniform}(\{1, \ldots,z\} \setminus \{x_i\})$.

\subsection{\label{sec:method_inference}Description of inference}

Compilation of new circuits using a trained model is straightforward.
Given the trained policy $\pi$, the starting state $s_0$ is constructed from the initial chip state and the initial circuit to be compiled.
Let $F_{\operatorname{total}}(\zeta)$ be the total duration under the trajectory $\zeta$.
The compilation now consists of sampling a trajectory $\zeta \sim \pi$ from the starting state $s_0$ under policy $\pi$.
Because this process is probabilistic, it can be repeated as often as desired and also be parallelized to achieve better results.
If repeated, the solution is that trajectory $\zeta$ with the smallest $F_{\operatorname{total}}(\zeta)$.
This approach has two advantages. 
First, it allows compilation in real time, since the next action can be sampled from the policy in constant time for any state.
More importantly, under an increasing time budget the compiler can generate increasingly efficient shuttling sequences (see Section~\ref{sec:TimeInference}).

\section{Results}
We train our proposed method RLIonS as described in Section~\ref{sec:proposed} using the hyperparameters from Appendix~\ref{sec:hyper} for up to 50 ions for each chip architecture: the X-chip, and two variants of the Q-chip as described in Section~\ref{sec:exp-architectures}. 
We do not tune hyperparameters per experiment such that the general applicability of our method can be evaluated.

\subsection{\label{sec:results_MQT}Main experiment on MQT bench circuits}
To compare the results from our new reinforcement learning method RLIonS with those of the exact SAT solver and the heuristic approach, we use the MQT bench dataset~\cite{mqt}. This dataset contains well-known quantum circuits, that are often used in other algorithms and serve as established benchmarks. It includes quantum circuits such as Grover's algorithm, the Quantum Fourier Transform (QFT), Quantum Approximation Optimization Algorithm (QAOA), and others. For each circuit, the number of qubits can be chosen. For our analysis, we use target-independent MQT circuits with optimization level 3, and include all circuits that we can transpile to our native gates, and for which at least the smallest instance is solvable with the SAT solver. As mentioned before, finding an exact solution breaks down when the problem dimension gets too big. Here, we terminated solver runs if the wall-time exceeds \SI{10}{\hour} for one boolean function $\Phi_T$. The maximum qubit number the SAT solver could handle varied between 3 and 20 qubits within the different problems of the MQT bench. In the majority of the cases, it stayed far below 20 qubits.  
We compare our proposed RLIonS compiler on the X-chip with the SAT solver and heuristic compiler.
While training for up to 30 ions would be sufficient for this experiment, we have used the same trained agent on the X-chip for all experiments to show that once trained RLIonS can solve a variety of problems.

\begin{figure}[h]
    \centering
    \includegraphics[width=1.0\linewidth]{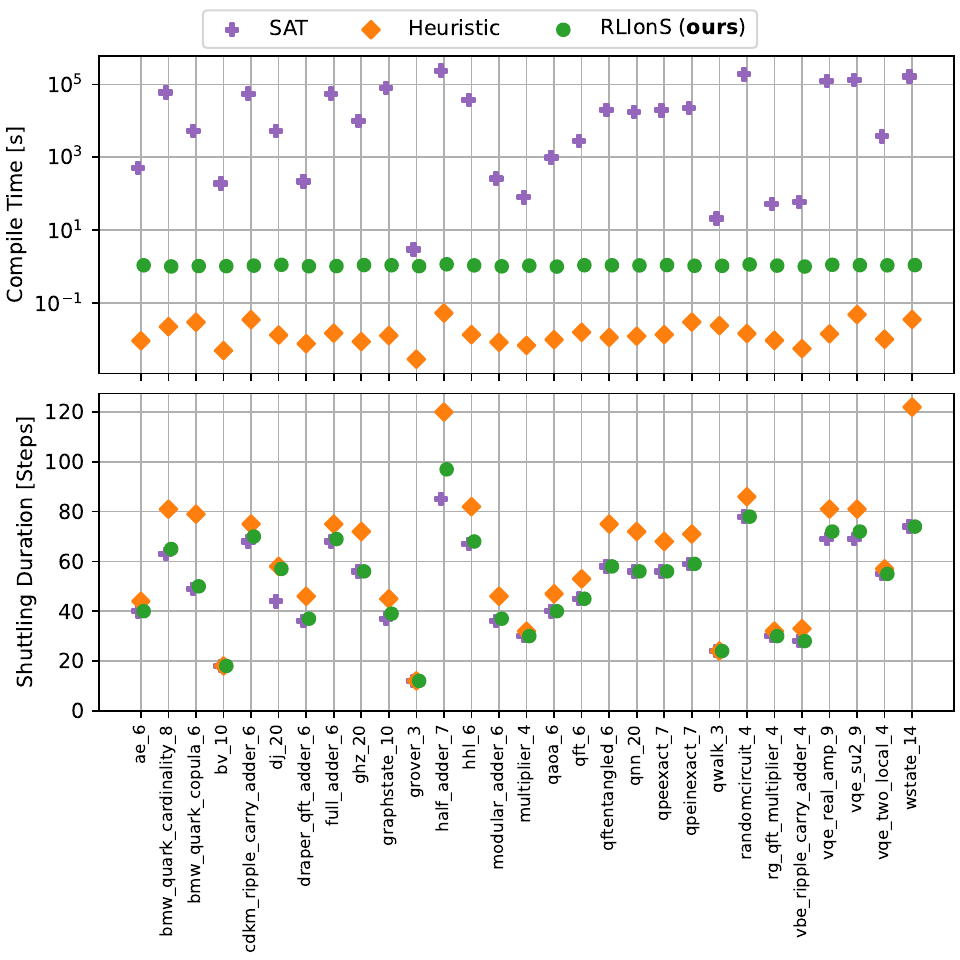}
    \caption{\label{fig:result-mqt-scatter-opt}Comparison of ion shuttling durations using trajectories optimized by different methods on the QVLS X-chip. For each type of problem within the MQT dataset, the largest example (the number at the end of the problem name indicates the qubit number) that could still be optimized by the SAT solver is shown. On average, the shuttling sequences generated by our proposed RL compiler are \SI{2.6}{\percent} worse than the optimal solutions of the SAT solver and \SI{14.0}{\percent} better than those of the heuristic compiler. The SAT solver requires \SI{12}{\hour} per problem on average, our approach \SI{1.1}{\second}.}
\end{figure}

For each circuit type, we select the instance with the largest number of ions for which the SAT solver was still able to find a solution when shuttling on the X-chip.
Figure~\ref{fig:result-mqt-scatter-opt} shows, for this set of circuits, the compile time (upper chart) and the resulting shuttling duration in steps (lower chart) for the three methods. RLIonS outperforms the heuristic compiler for all circuits and has, for most circuits, a similar or only slightly worse performance than the exact SAT solver. On average, of the circuits that were solvable for the SAT solver, RLIonS performs only $\SI{2.6}{\percent}$ worse than optimum. To put this result into context, it is important to look at the compile time: While RLIonS ($\SI{1.1}{\second}$) and the heuristic compiler ($<\SI{0.1}{\second}$) are very fast, the SAT solver takes on average $\SI{12}{\hour}$ to find its exact solution. This timescale is not feasible for actual application.
Furthermore, as explained in Section~\ref{sec:TimeInference}, the optimality gap can be further closed by increasing RLIonS' compile time slightly, while still being much faster than the SAT solver and remaining within an acceptable compile-time regime for the actual use case.

For further comparison, in Table~\ref{tab:result-mqt-largest}, we select the largest version of each circuit type in our dataset and compare RLIonS to the heuristic compiler. Both methods find shuttling schedules for all of these problems. On average, RLIonS takes \SI{19.8}{\percent} fewer steps than the heuristic compiler.

Additionally, the results for all studied MQT circuits as well as animations of some example shuttling sequences are presented in the Supplementary Material~\cite{supp_material}.

A common practice in RL is to evaluate on the same domain and distribution that were used for training.  A good shuttling compiler, however, should perform well on arbitrary circuits. Because our training circuits are generated by uniformly sampling random gates, complex MQT circuits with specific gate patterns were unlikely to appear during training. It is therefore a crucial finding that our agent generalizes effectively to previously unseen problems.

\begin{table}[]
    \centering
    \footnotesize
    \begin{tabular}{lrr}
\toprule
\thead{Problem} & \thead{RLIonS\\Steps} & \thead{Heuristic\\Steps} \\
\midrule
\texttt{ae\_15} & 319 & 356 \\
\texttt{bmw\_quark\_cardinality\_20} & 186 & 242 \\
\texttt{bmw\_quark\_copula\_20} & 678 & 845 \\
\texttt{bv\_10} & 18 & 18 \\
\texttt{cdkm\_ripple\_carry\_adder\_20} & 309 & 332 \\
\texttt{dj\_20} & 57 & 58 \\
\texttt{draper\_qft\_adder\_20} & 493 & 668 \\
\texttt{full\_adder\_20} & 309 & 332 \\
\texttt{ghz\_32} & 92 & 120 \\
\texttt{graphstate\_20} & 107 & 149 \\
\texttt{grover\_7} & 3081 & 3499 \\
\texttt{half\_adder\_19} & 744 & 1118 \\
\texttt{hhl\_20} & 1179 & 1519 \\
\texttt{hrs\_cumulative\_multiplier\_17} & 4122 & 4625 \\
\texttt{modular\_adder\_20} & 493 & 668 \\
\texttt{multiplier\_20} & 2879 & 4675 \\
\texttt{qaoa\_20} & 835 & 1167 \\
\texttt{qft\_20} & 609 & 715 \\
\texttt{qftentangled\_30} & 1448 & 1855 \\
\texttt{qnn\_20} & 56 & 72 \\
\texttt{qpeexact\_20} & 577 & 700 \\
\texttt{qpeinexact\_20} & 585 & 700 \\
\texttt{qwalk\_7} & 3377 & 3957 \\
\texttt{randomcircuit\_20} & 4355 & 5681 \\
\texttt{rg\_qft\_multiplier\_20} & 2914 & 4675 \\
\texttt{vbe\_ripple\_carry\_adder\_19} & 361 & 448 \\
\texttt{vqe\_real\_amp\_20} & 186 & 246 \\
\texttt{vqe\_su2\_20} & 186 & 246 \\
\texttt{vqe\_two\_local\_20} & 1898 & 2622 \\
\texttt{wstate\_20} & 110 & 122 \\
\bottomrule
\end{tabular}
    \caption{\label{tab:result-mqt-largest}Comparison of ion shuttling durations using trajectories optimized by the heuristic compiler and our proposed method on the X-chip for the largest version per type of problem within our dataset of the MQT circuits. On average, our proposed compiler generates \SI{19.8}{\percent} faster shuttling sequences.}
\end{table}

\subsection{Main experiment on quantum volume circuits}
Another widely used benchmark are the quantum volume (QV) circuits, which were developed to compare the quality of quantum computers across different platforms~\cite{crossValidatingQuantumComputers2019}. They are random square-shaped circuits, meaning the number of qubits equals the circuit depth. In each layer, the qubits undergo a random permutation after which two-qubit gates are applied to neighboring qubits. As these circuits have a well-defined structure and can be generated for any number of qubits, we use them to evaluate RLIonS' performance for increasing qubit counts. 
For this experiment, we compare our RLIonS compiler with the heuristic compiler on the X-chip. The SAT compiler is not included here as it does not scale beyond 6 qubits, see section~\ref{sec:TimeInference} for a detailed comparison on QV(6) circuits.

\begin{figure}[h]
    \centering
    \includegraphics[width=1.0\linewidth]{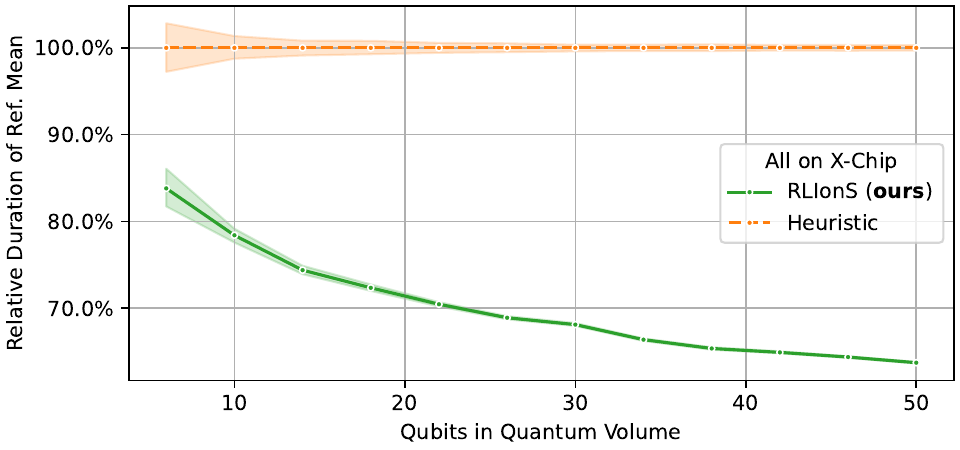}
    \caption{\label{fig:result-qv-comp}Comparison of ion shuttling duration of our proposed method to the reference heuristic compiler on the QVLS X-chip for increasingly difficult quantum volume circuits. 
    Shaded areas indicate \SI{95}{\percent} confidence intervals.
    With a comparable inference time budget, our proposed compiler finds significantly more efficient ion shuttling trajectories. The advantage of our approach increases from \SI{16.2}{\percent} on small circuits up to \SI{36.3}{\percent} on the largest ones.}
\end{figure}

Figure~\ref{fig:result-qv-comp} shows RLIonS' resulting shuttling duration relative to the duration obtained by the heuristic compiler on the X-chip for qubit numbers between 6 and 50.
For comparable results, the compilation with RLIonS is stopped if it exceeds the maximum of the compile time of the heuristic compiler and \SI{1}{\second}.
RLIonS achieves much faster shuttling schedules, yielding an advantage of up to $\SI{36.3}{\percent}$ for large problems that require more than $10.000$ shuttling steps. 
It is especially important that the speed-up the reinforcement learning method provides increases with increasing ion number.
This makes our method particularly useful, as the problem of executing long quantum circuits within the ions' coherence time becomes more pronounced as qubit numbers increase.

\subsection{\label{sec:TimeInference}Scaling with time during inference}

In Section~\ref{sec:method_inference} we have outlined how the probabilistic rollout during compilation can be used to generate increasingly good solutions with increasing time. 
We have analyzed this scaling on quantum volume problems with 6 qubits in Table~\ref{tab:results-sat-scaling}. 
Our proposed method is simulated with a time budget scaling from \SI{100}{\milli\second} up to \SI{100}{\second}. 
This is done by compiling \SI{100}{} predetermined QV(6) circuits for \SI{600}{\second} per circuit with RLIonS on the X-chip.
Then, bootstrapping is used to draw \SI{20000}{} samples of the respective budget, e.g. \SI{100}{\milli\second}, to determine the average metrics reported in the table. A comparison with the heuristic compiler and SAT solver on the same \SI{100}{} circuits is included. The optimality gap is reported, which is the difference in normalized steps between the proposed solution and the optimal solution. 
For each ten-fold increase of inference time, our method can close the average optimality gap further, going from a difference of \SI{0.71}{steps} for \SI{0.1}{\second} compilation time to only \SI{0.42}{steps} for \SI{100}{\second} compilation time.
These findings highlight the versatility of our proposed approach, initially finding already very good solutions compared to the heuristic compiler, which can further be improved with moderate increases in compile time compared to the SAT solver.

\begin{table*}[]
    \centering
\begin{tabular}{l|r|rrrr|r}
\toprule
\multirow{2}{*}{\thead{Method}} & \thead{Compile time} & \multicolumn{4}{r|}{\thead{Optimality gap}} & \thead{Average}  \\
 & \thead{per problem} & \thead{0 steps} & \thead{1 step} & \thead{2 steps} & \thead{$>$ 2 steps} & \thead{opt. gap} \\
\midrule
RLIonS (ours) & \SI{0.1}{\second} & \SI{47.24}{\percent} & \SI{36.28}{\percent} & \SI{14.66}{\percent} & \SI{1.82}{\percent} & \SI{0.71}{steps} \\
RLIonS (ours) & \SI{1.0}{\second} & \SI{51.28}{\percent} & \SI{36.36}{\percent} & \SI{12.32}{\percent} & \SI{0.04}{\percent} & \SI{0.61}{steps} \\
RLIonS (ours) & \SI{10.0}{\second} & \SI{57.52}{\percent} & \SI{34.08}{\percent} & \SI{8.40}{\percent} & \SI{0.00}{\percent} & \SI{0.51}{steps} \\
RLIonS (ours) & \SI{100.0}{\second} & \SI{62.32}{\percent} & \SI{33.08}{\percent} & \SI{4.60}{\percent} & \SI{0.00}{\percent} & \SI{0.42}{steps} \\
\midrule
Heuristic & \SI{0.1}{\second} & \SI{0.00}{\percent} & \SI{0.00}{\percent} & \SI{0.00}{\percent} & \SI{100.00}{\percent} & \SI{10.65}{steps} \\
SAT & \SI{12766.5}{\second} & \SI{100.00}{\percent} & \SI{0.00}{\percent} & \SI{0.00}{\percent} & \SI{0.00}{\percent} & \SI{0.00}{steps} \\
\bottomrule
\end{tabular}
    \caption{\label{tab:results-sat-scaling}Optimality gap (absolute difference in steps to the optimal solution) on 100 randomly generated quantum volume problems with 6 qubits. The optimal solutions were computed using a SAT solver and have an average of $48.52$ steps. 
    With a larger inference time budget, the average optimality gap of our method closes and our method is more likely to find the optimal solution. 
    With a time budget of \SI{10}{\second}, RLIonS always proposes solutions within 2 step of the optimum. As the SAT solver is the ground truth, it is always optimal. However, it requires more than \SI{3}{\hour} per problem on average for these relatively small circuits.
    The heuristic compiler is as fast as when running our method with the lowest time budget, but has a larger optimality gap than our proposed method and is never within 2 steps of the optimal solution.}
\end{table*}

\subsection{\label{sec:exp-architectures}Extension to other architectures}
So far, our proposed RL ion shuttling approach has been applied to a chip with an X-junction and two separate storage stacks. We now evaluate how well our method generalizes to other architectures. We focus on a Q-chip design as shown in Figure~\ref{fig:q-chip} and discussed in Section~\ref{sec:Q-chip}. 
It is difficult to know a priori how different design choices will affect shuttling efficiency. It is therefore invaluable to test architectures using an actual shuttling compiler to see how they perform on realistic tasks.

We train RLIonS on a Q-chip with a capacity of up to 50 qubits in the same fashion as RLIonS was trained on the X-chip. We also train on a modified Q-chip, increasing the capacity of the SPAM zone from 1 to 3 qubits. As explained in Section~\ref{sec:Q-chip}, we expect that the duration of one shuttling step is shorter in this architecture when empty spots cross the junction. The actual time difference has to be determined experimentally after the chip is fabricated. Because the SAT solver only supports uniform step sizes, it is not applicable here. A heuristic compiler does not currently exist for this architecture. 

\begin{figure}
    \centering
    \includegraphics[width=1.0\linewidth]{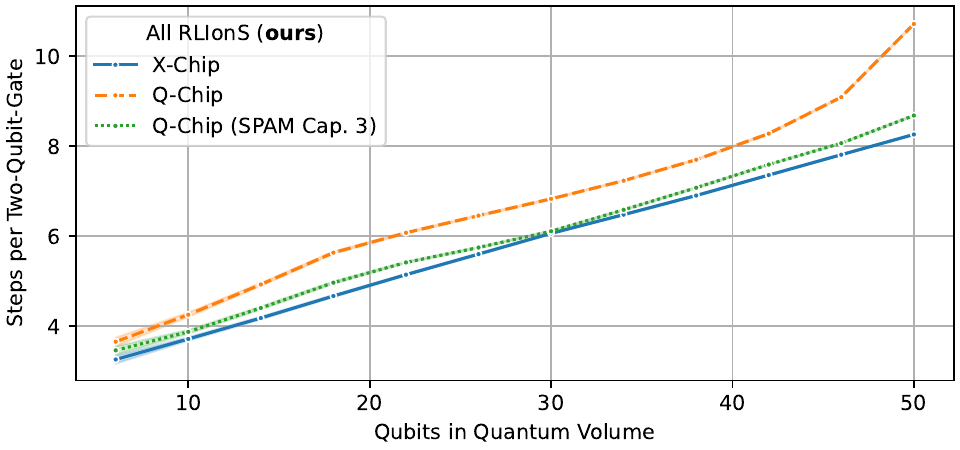}
    \caption{\label{fig:result-architectures}Ion shuttling duration for different architectures optimized with our proposed RL method.
    Shaded areas indicate \SI{95}{\percent} confidence intervals.
    All tested architectures are consistently optimized by our method.
    The generated shuttling sequences for the standard Q-chip use significantly more steps per two-qubit gate than the X-chip for all tested problem sizes. This can be mitigated by increasing the capacity of the SPAM zone from 1 to 3 qubits. 
    Such exploratory design studies are very easy to perform using our method, as it does not require the development of new heuristics or other rules for new hardware.}
\end{figure}

We report RLIonS’ performance on quantum volume circuits in Figure~\ref{fig:result-architectures}. Ion shuttling on the standard Q-chip is consistently less efficient than on the X-chip: for up to 40 ions, it incurs roughly one extra shuttling step per two-qubit gate, and the overhead grows sharply beyond that. This abrupt rise can be explained by the storage ring reaching capacity, limiting opportunities to insert ions from other registers.

When raising the capacity of the SPAM register from 1 to 3 ions, the Q-chip performs almost as efficiently as the X-chip. This increase in intermediate storage options, which is likely achievable with no significant hardware changes, would therefore be a helpful mitigation design choice.

This analysis shows that our RL method is highly versatile and can successfully optimize ion shuttling across different architectures. We also demonstrate that it can be used to explore different hardware decisions during the design phase.

\subsection{\label{sec:ablations}Ablations: Analysis of RL design choices}

Our proposed method contains various design decisions. In this section we briefly evaluate the importance of some key components. The following ablations are considered: 1) Using a linear encoding of numeric values in the representation instead of the sinusoidal encoding described in Appendix~\ref{sec:appendix_sinusoidal}. 2) Setting $\gamma = \gamma_s$, such that the shaped reward from section~\ref{sec:shaped_reward} does not introduce greediness. 3) Disabling the shaped reward completely. 4) Selecting a more naive representation (see Appendix~\ref{sec:appendix_naive}) than our proposed representation from Figure~\ref{fig:representation} and Section~\ref{sec:proposed_representation}.

\begin{table}[]
    \centering
\begin{tabular}{l|r}
\toprule
\thead{Method} & \thead{Largest solved \\quantum volume circuit} \\
\midrule
RLIonS & \SI{50}{qubits} \\
RLIonS (Linear numeric encoding) & \SI{50}{qubits} \\
RLIonS ($\gamma = \gamma_s$) & \SI{26}{qubits} \\
RLIonS (No shaped reward) & \SI{26}{qubits} \\
RLIonS (Naive representation) & \SI{0}{qubits} \\
\bottomrule
\end{tabular}
    \caption{\label{tab:ablations}Ablation study on the importance of individual components of RLIonS.
    While all proposed design choices are important, our proposed representation has the largest effect. When using a more naive representation that is not invariant to qubit relabeling or commutative circuit reordering, the RL approach fails to solve even the smallest circuits.}
\end{table}

\begin{figure}
    \centering
    \includegraphics[width=1.0\linewidth]{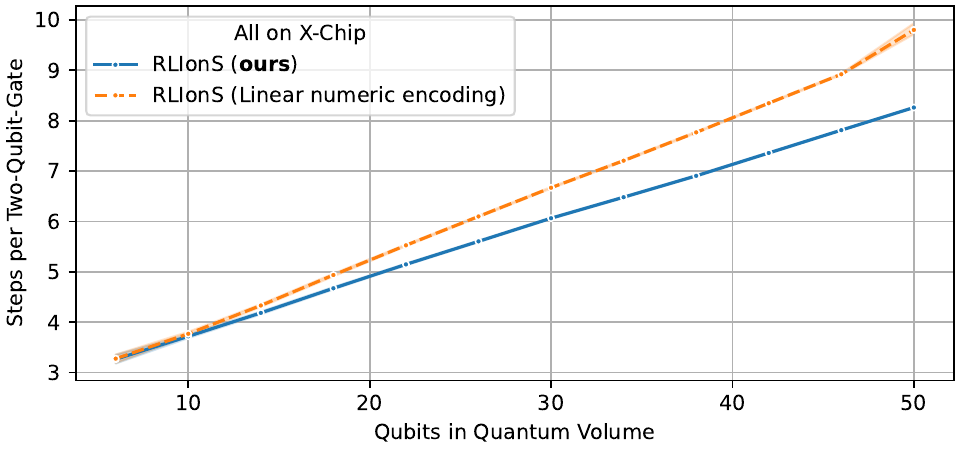}%
    \caption{\label{fig:ablations}
    Detailed study on the influence of numeric encodings in RLIonS, which is the only ablation experiment where compilation remains reliable up to 50 qubits.
    When using a linear encoding instead of the proposed sinusoidal encoding, the quality of shuttling sequences is reduced for problems larger than 10 qubits.
}
\end{figure}

An agent is trained for each ablation on the X-chip with a capacity of 50 qubits and compared to the unmodified proposed method on quantum volume circuits. In Table~\ref{tab:ablations}, the largest size of quantum volume circuits that can be solved is reported. A QV size is considered solvable if the agent reliably finds a valid shuttling sequence for 100 of 100 test cases.
For most ablations, the largest solvable size is severely reduced.
A greedy shaped reward is important, as without greediness, only up to 26 qubits can be solved. Without a shaped reward at all, again only 26 qubits can be solved.
Our proposed representation has the largest influence by far. With a naive representation, the RL approach fails to compile any QV circuits, as it fails to learn the underlying patterns of the representation.
When using a linear numeric encoding instead of the sinusoidal encoding, compilation remains reliable. We therefore 
briefly compare the quality of the compilations of both methods in Figure~\ref{fig:ablations}.
The agent using the linear encoding produces longer shuttling sequences for the same problems.

\section{Conclusion and outlook}

We show that reinforcement learning (RL) can produce efficient ion‑shuttling schedules for trapped‑ion quantum computers based on the quantum charge‑coupled device (QCCD) architecture. Our schedules need up to $\SI{36.3}{\percent}$ fewer shuttling operations than state‑of‑the‑art heuristic compilers. 
As shuttling dominates circuit runtime, the speed-up substantially enhances the reliability of trapped‑ion quantum computers since they are constrained by finite coherence times.
For small instances, we benchmark our method, RLIonS, against exact SAT solutions. In most cases, RLIonS performs near optimally. For larger problems, exact solutions are intractable.

We also showcase the versatility of our approach, successfully applying it across different architectures. This makes our method useful during hardware development to assess how design choices affect shuttling efficiency.

Future work includes extending to more complex hardware, e.g., multi‑junction chips. As complexity increases, developing efficient heuristic strategies will become even more difficult. Our observation encoding could be generalized to further reduce shuttling overhead for circuits with long chains of commuting two-qubit gates. RL could also be used to optimize the initial qubit‑to‑ion mapping, to evaluate design choices such as ion swapping or chain splitting, and to optimize architectures that support parallel operations, such as simultaneous storage rotations and swaps in another register.

Overall, our results underscore the importance of integrating methods from physics and computer science to advance trapped‑ion quantum computing.

\begin{acknowledgments}

We thank Celeste Torkzaban and Florian Ungerechts for the invaluable exchange on the current and planned chip designs at QVLS.
This work was funded by the Deutsche Forschungsgemeinschaft (DFG, German Research Foundation) under Germany’s Excellence Strategy –
EXC-2123 QuantumFrontiers – 390837967, the state of Lower Saxony and the VW foundation through Quantum Valley Lower Saxony Q1 (QVLS-Q1), the Federal Ministry of Research, Technology and
Space (BMFTR) within the Quantum Computing Service Center QUICS (grant no. 13N17418) and  QuSTAC (FKZ 13N17319).

\end{acknowledgments}

\appendix
\section{\label{sec:hyper}Hyperparameters}

See Table~\ref{tab:hparams} for a list of the used values of all hyperparameters. 
These hyperparameters were determined empirically, with the exception of the step durations for the Q-chip. Here, the fast rotation speed-up is an estimated value. The final value on the real hardware is currently not known.

\begin{table}[h!]
    \centering
    \begin{tabular}{lr}
        \toprule
        \thead{Hyperparameter} & \thead{Value} \\
        \midrule
        Initial learning rate $\alpha$ & $2.5 \cdot 10^{-4}$ \\
        Learning rate decay & Linear \\
        Value function coefficient & 0.5 \\
        Clipping factor $\epsilon$ & $0.1$ \\
        Parallel environments $n_{\operatorname{envs}}$ & $250$ \\
        Steps per environment $n_{\operatorname{steps}}$ & $40$ \\
        Entropy coefficient & $1.0 \cdot 10^{-4}$ \\
        Maximum epochs per learning step & $4$ \\
        Mini batch size & $1024$ \\
        GAE $\lambda$ & $0.96$ \\
        Discount factor $\gamma$ ($= e^{-\beta}$) & $0.9995$ \\
        Shaped reward discount factor $\gamma_s$ & 1.0 \\
        Penalty rate $c_r$ & 0.1 \\
        Training gate count $n_{\operatorname{gates}}$ & 1275 \\
        Residual network blocks $n_{\operatorname{blocks}}$ & 3 \\
        Hidden layer width $n_{\operatorname{hidden}}$ & 512 \\
        Shared value/policy network & No \\
        PPO learning steps & \SI{1000000}{} \\
        \midrule
        Gate depth in representation $k_{\text{lookahead}}$ & 4 \\
        Sinusoidal bands cell numbers $b_{\operatorname{cell}}$ & 6 \\
        Sinusoidal bands remaining total gates $b_{\operatorname{total\_gates}}$ & 7 \\
        \midrule
        X-chip movement duration & \SI{1.00}{steps} \\
        Q-chip default movement duration & \SI{1.00}{steps} \\
        Q-chip fast rotation duration & \SI{0.25}{steps} \\
         \bottomrule
    \end{tabular}
    \caption{\label{tab:hparams}Hyperparameters used for the PPO algorithm and the ResNet architecture, our proposed representation, and the Q-chip.}
\end{table}

\section{Hardware requirements and libraries}
The SAT solver uses the Z3-solver~\cite{de2008z3} running on an Intel(R) Xeon(R) Gold 6242R CPU @ 3.10GHz. The proposed RL approach is implemented using the JAX high performance numerical computing library, running on an Intel(R) Core(TM) i9-9900K CPU @ 3.60GHz and an NVIDIA GeForce RTX 2080 Ti.

\section{\label{sec:appendix_sinusoidal}Sinusoidal numeric embeddings}
We employ a sinusoidal embedding~\cite{vaswani2017attention} to encode the numeric values of the representation, such as cell numbers. This embedding is a parameterized function $S(x; x_{\operatorname{max}}, b)$ with the number of bands $b$ and the maximum encoded value $x_{\operatorname{max}}$. These parameters are not optimized through gradient descent. $S$ is defined as:
\begin{align}
    S(x;x_{\operatorname{max}}, b) &= \begin{pmatrix}
        x_{\operatorname{norm}} \\
        \operatorname{cos}(x_{\operatorname{norm}} \cdot \pi \cdot 2^0) \\
        \ldots \\
        \operatorname{cos}(x_{\operatorname{norm}} \cdot \pi \cdot 2^{b-1}) \\
        \operatorname{sin}(x_{\operatorname{norm}} \cdot \pi \cdot 2^0) \\
        \ldots \\
        \operatorname{sin}(x_{\operatorname{norm}} \cdot \pi \cdot 2^{b-1}) 
    \end{pmatrix} \, \text{, where} \\
    x_{\operatorname{norm}} &= \operatorname{clip}\left(\frac{x}{x_{\operatorname{max}}}, 0, 1\right) \, .
\end{align}
A special case arises if $x$ does not represent a valid number, i.e. when encoding the value $\diamond$ signaling empty. In this case, $S(\diamond; x_{\operatorname{max}}, b) = \mathbf{0}$.

\section{\label{sec:appendix_naive}Naive representation}
In section~\ref{sec:ablations} we use a more naive representation to highlight the importance of our proposed representation. The naive representation encodes the qubit positions as a vector of cells $\mathbf{v}_q$, where each entry is either the qubit label or zero if empty. For the example in Figure~\ref{fig:representation} this would be $\mathbf{v}_q = (4, 1, 0, 0, 3, 5, 0, 0, 0, 2, 0)$. This is concatenated with a vector $\mathbf{v}_g$ describing the circuit by listing the remaining gates directly through the operand labels. This is padded with zeros to a fixed size. The circuit in Figure~\ref{fig:representation} would be $\mathbf{v}_g = (1, 3, 2, 4, 1, 5, 1, 3, 0, 0, \ldots)$. Importantly, this representation contains at least the same amount of information as our proposed representation, but does not alias equivalent states.

\bibliography{rl_ion_shuttling,Bib_Lea}

\end{document}


\title{Supplementary Material: Reinforcement learning for ion shuttling on trapped-ion quantum computers}

\author{Maximilian Schier}
\email[Equal contribution, order decided randomly. Contact authors: ]{schier@tnt.uni-hannover.de, lea.richtmann@aei.uni-hannover.de}
\affiliation{Institute for Information Processing (tnt), L3S, Leibniz University Hannover, Germany}
\author{Lea Richtmann}
\email[Equal contribution, order decided randomly. Contact authors: ]{schier@tnt.uni-hannover.de, lea.richtmann@aei.uni-hannover.de}
\affiliation{Institute for Gravitational Physics, Leibniz University Hannover, Germany}
\author{Christian Staufenbiel}
\affiliation{QUDORA Technologies GmbH}
\author{Tobias Schmale}
\affiliation{QUDORA Technologies GmbH}
\affiliation{Institute for Theoretical Physics, Leibniz University Hannover, Germany}
\author{Daniel Borcherding}
\affiliation{QUDORA Technologies GmbH}
\author{Michèle Heurs}
\affiliation{Institute for Gravitational Physics, Leibniz University Hannover, Germany}
\affiliation{Deutsches Zentrum für Astrophysik (DZA)}
\affiliation{Deutsches Elektronen-Synchrotron (DESY)}
\author{Bodo Rosenhahn}
\affiliation{Institute for Information Processing (tnt), L3S, Leibniz University Hannover, Germany}
\maketitle

\section{Full results on MQT circuits}
The following table lists all results of the SAT solver, our proposed approach RLIonS, and the heuristic compiler on the X-chip. A dash in the column of the SAT solver means that it failed to compute a shuttling schedule for this problem.
{
\begin{longtblr}[
  caption = {Full results on MQT circuits},
  label = {tab:sup-mqt-results},
]{
  colspec = {lrrr},
  rowhead = 1, 
  rowsep = 0.5pt, 
  hline{1,Z} = {0.08em}, 
  hline{2} = {0.05em},   
}
Problem & SAT steps & RLIonS steps & Heuristic steps \\
\texttt{ae\_3} & 8 & 8 & 10 \\
\texttt{ae\_4} & 16 & 16 & 16 \\
\texttt{ae\_5} & 26 & 26 & 30 \\
\texttt{ae\_6} & 40 & 40 & 44 \\
\texttt{ae\_7} & - & 58 & 65 \\
\texttt{ae\_8} & - & 79 & 94 \\
\texttt{ae\_9} & - & 102 & 115 \\
\texttt{ae\_10} & - & 134 & 153 \\
\texttt{ae\_11} & - & 160 & 219 \\
\texttt{ae\_12} & - & 195 & 238 \\
\texttt{ae\_13} & - & 232 & 264 \\
\texttt{ae\_14} & - & 273 & 374 \\
\texttt{ae\_15} & - & 319 & 356 \\
\texttt{bmw\_quark\_cardinality\_3} & 14 & 14 & 14 \\
\texttt{bmw\_quark\_cardinality\_4} & 26 & 26 & 29 \\
\texttt{bmw\_quark\_cardinality\_5} & 34 & 34 & 44 \\
\texttt{bmw\_quark\_cardinality\_6} & 44 & 45 & 53 \\
\texttt{bmw\_quark\_cardinality\_7} & 53 & 54 & 63 \\
\texttt{bmw\_quark\_cardinality\_8} & 63 & 65 & 81 \\
\texttt{bmw\_quark\_cardinality\_9} & - & 77 & 91 \\
\texttt{bmw\_quark\_cardinality\_10} & - & 84 & 108 \\
\texttt{bmw\_quark\_cardinality\_11} & - & 94 & 119 \\
\texttt{bmw\_quark\_cardinality\_12} & - & 104 & 136 \\
\texttt{bmw\_quark\_cardinality\_13} & - & 114 & 146 \\
\texttt{bmw\_quark\_cardinality\_14} & - & 125 & 161 \\
\texttt{bmw\_quark\_cardinality\_15} & - & 136 & 180 \\
\texttt{bmw\_quark\_cardinality\_16} & - & 149 & 188 \\
\texttt{bmw\_quark\_cardinality\_17} & - & 157 & 198 \\
\texttt{bmw\_quark\_cardinality\_18} & - & 168 & 213 \\
\texttt{bmw\_quark\_cardinality\_19} & - & 175 & 242 \\
\texttt{bmw\_quark\_cardinality\_20} & - & 186 & 242 \\
\texttt{bmw\_quark\_copula\_4} & 12 & 12 & 14 \\
\texttt{bmw\_quark\_copula\_6} & 49 & 50 & 79 \\
\texttt{bmw\_quark\_copula\_8} & - & 92 & 94 \\
\texttt{bmw\_quark\_copula\_10} & - & 149 & 180 \\
\texttt{bmw\_quark\_copula\_12} & - & 218 & 312 \\
\texttt{bmw\_quark\_copula\_14} & - & 316 & 357 \\
\texttt{bmw\_quark\_copula\_16} & - & 426 & 540 \\
\texttt{bmw\_quark\_copula\_18} & - & 519 & 751 \\
\texttt{bmw\_quark\_copula\_20} & - & 678 & 845 \\
\texttt{bv\_3} & 4 & 4 & 4 \\
\texttt{bv\_4} & 6 & 6 & 6 \\
\texttt{bv\_5} & 8 & 8 & 8 \\
\texttt{bv\_6} & 10 & 10 & 10 \\
\texttt{bv\_7} & 12 & 12 & 12 \\
\texttt{bv\_8} & 14 & 14 & 14 \\
\texttt{bv\_9} & 16 & 16 & 16 \\
\texttt{bv\_10} & 18 & 18 & 18 \\
\texttt{cdkm\_ripple\_carry\_adder\_4} & 36 & 36 & 36 \\
\texttt{cdkm\_ripple\_carry\_adder\_6} & 68 & 70 & 75 \\
\texttt{cdkm\_ripple\_carry\_adder\_8} & - & 104 & 111 \\
\texttt{cdkm\_ripple\_carry\_adder\_10} & - & 138 & 144 \\
\texttt{cdkm\_ripple\_carry\_adder\_12} & - & 171 & 183 \\
\texttt{cdkm\_ripple\_carry\_adder\_14} & - & 205 & 222 \\
\texttt{cdkm\_ripple\_carry\_adder\_16} & - & 239 & 260 \\
\texttt{cdkm\_ripple\_carry\_adder\_18} & - & 273 & 291 \\
\texttt{cdkm\_ripple\_carry\_adder\_20} & - & 309 & 332 \\
\texttt{dj\_3} & 8 & 8 & 10 \\
\texttt{dj\_4} & 10 & 10 & 10 \\
\texttt{dj\_5} & 12 & 12 & 13 \\
\texttt{dj\_6} & 15 & 15 & 16 \\
\texttt{dj\_7} & 18 & 18 & 19 \\
\texttt{dj\_8} & 20 & 21 & 22 \\
\texttt{dj\_9} & 22 & 24 & 25 \\
\texttt{dj\_10} & 24 & 27 & 28 \\
\texttt{dj\_11} & 26 & 30 & 31 \\
\texttt{dj\_12} & 28 & 33 & 34 \\
\texttt{dj\_13} & 30 & 36 & 37 \\
\texttt{dj\_14} & 32 & 39 & 40 \\
\texttt{dj\_15} & 34 & 42 & 43 \\
\texttt{dj\_16} & 36 & 45 & 46 \\
\texttt{dj\_17} & 38 & 48 & 49 \\
\texttt{dj\_18} & 40 & 51 & 52 \\
\texttt{dj\_19} & - & 54 & 55 \\
\texttt{dj\_20} & 44 & 57 & 58 \\
\texttt{draper\_qft\_adder\_4} & 17 & 17 & 23 \\
\texttt{draper\_qft\_adder\_6} & 36 & 37 & 46 \\
\texttt{draper\_qft\_adder\_8} & - & 69 & 83 \\
\texttt{draper\_qft\_adder\_10} & - & 111 & 131 \\
\texttt{draper\_qft\_adder\_12} & - & 165 & 239 \\
\texttt{draper\_qft\_adder\_14} & - & 223 & 290 \\
\texttt{draper\_qft\_adder\_16} & - & 302 & 398 \\
\texttt{draper\_qft\_adder\_18} & - & 397 & 521 \\
\texttt{draper\_qft\_adder\_20} & - & 493 & 668 \\
\texttt{full\_adder\_4} & 28 & 28 & 33 \\
\texttt{full\_adder\_6} & 68 & 69 & 75 \\
\texttt{full\_adder\_8} & - & 104 & 111 \\
\texttt{full\_adder\_10} & - & 137 & 144 \\
\texttt{full\_adder\_12} & - & 172 & 183 \\
\texttt{full\_adder\_14} & - & 205 & 222 \\
\texttt{full\_adder\_16} & - & 239 & 260 \\
\texttt{full\_adder\_18} & - & 274 & 291 \\
\texttt{full\_adder\_20} & - & 309 & 332 \\
\texttt{ghz\_3} & 4 & 4 & 4 \\
\texttt{ghz\_4} & 8 & 8 & 8 \\
\texttt{ghz\_5} & 10 & 10 & 12 \\
\texttt{ghz\_6} & 14 & 14 & 16 \\
\texttt{ghz\_7} & 16 & 16 & 20 \\
\texttt{ghz\_8} & 20 & 20 & 24 \\
\texttt{ghz\_9} & 22 & 22 & 28 \\
\texttt{ghz\_10} & 26 & 26 & 32 \\
\texttt{ghz\_11} & 28 & 28 & 36 \\
\texttt{ghz\_12} & 32 & 32 & 40 \\
\texttt{ghz\_13} & 34 & 34 & 44 \\
\texttt{ghz\_14} & 38 & 38 & 48 \\
\texttt{ghz\_15} & 40 & 40 & 52 \\
\texttt{ghz\_16} & 44 & 44 & 56 \\
\texttt{ghz\_17} & 46 & 46 & 60 \\
\texttt{ghz\_18} & 50 & 50 & 64 \\
\texttt{ghz\_19} & 52 & 52 & 68 \\
\texttt{ghz\_20} & 56 & 56 & 72 \\
\texttt{ghz\_30} & - & 86 & 112 \\
\texttt{ghz\_32} & - & 92 & 120 \\
\texttt{graphstate\_3} & 10 & 10 & 12 \\
\texttt{graphstate\_4} & 14 & 14 & 16 \\
\texttt{graphstate\_5} & 17 & 17 & 21 \\
\texttt{graphstate\_6} & 22 & 22 & 26 \\
\texttt{graphstate\_7} & 25 & 25 & 33 \\
\texttt{graphstate\_8} & 28 & 28 & 35 \\
\texttt{graphstate\_9} & 33 & 33 & 39 \\
\texttt{graphstate\_10} & 37 & 39 & 45 \\
\texttt{graphstate\_11} & - & 46 & 51 \\
\texttt{graphstate\_12} & - & 49 & 58 \\
\texttt{graphstate\_13} & - & 48 & 53 \\
\texttt{graphstate\_14} & - & 54 & 65 \\
\texttt{graphstate\_15} & - & 69 & 96 \\
\texttt{graphstate\_16} & - & 70 & 97 \\
\texttt{graphstate\_17} & - & 72 & 88 \\
\texttt{graphstate\_18} & - & 78 & 103 \\
\texttt{graphstate\_19} & - & 85 & 123 \\
\texttt{graphstate\_20} & - & 107 & 149 \\
\texttt{grover\_3} & 12 & 12 & 12 \\
\texttt{grover\_4} & - & 118 & 128 \\
\texttt{grover\_5} & - & 388 & 437 \\
\texttt{grover\_6} & - & 1032 & 1173 \\
\texttt{grover\_7} & - & 3081 & 3499 \\
\texttt{half\_adder\_3} & 12 & 12 & 12 \\
\texttt{half\_adder\_5} & 46 & 46 & 53 \\
\texttt{half\_adder\_7} & 85 & 97 & 120 \\
\texttt{half\_adder\_9} & - & 167 & 194 \\
\texttt{half\_adder\_11} & - & 258 & 338 \\
\texttt{half\_adder\_13} & - & 321 & 463 \\
\texttt{half\_adder\_15} & - & 437 & 642 \\
\texttt{half\_adder\_17} & - & 510 & 739 \\
\texttt{half\_adder\_19} & - & 744 & 1118 \\
\texttt{hhl\_3} & 6 & 6 & 6 \\
\texttt{hhl\_4} & 23 & 23 & 32 \\
\texttt{hhl\_5} & 42 & 43 & 46 \\
\texttt{hhl\_6} & 67 & 68 & 82 \\
\texttt{hhl\_7} & - & 102 & 123 \\
\texttt{hhl\_8} & - & 144 & 163 \\
\texttt{hhl\_9} & - & 186 & 233 \\
\texttt{hhl\_10} & - & 241 & 303 \\
\texttt{hhl\_11} & - & 299 & 386 \\
\texttt{hhl\_12} & - & 382 & 451 \\
\texttt{hhl\_13} & - & 450 & 560 \\
\texttt{hhl\_14} & - & 545 & 666 \\
\texttt{hhl\_15} & - & 632 & 845 \\
\texttt{hhl\_16} & - & 725 & 898 \\
\texttt{hhl\_17} & - & 838 & 1162 \\
\texttt{hhl\_18} & - & 952 & 1293 \\
\texttt{hhl\_19} & - & 1075 & 1400 \\
\texttt{hhl\_20} & - & 1179 & 1519 \\
\texttt{hrs\_cumulative\_multiplier\_5} & - & 263 & 288 \\
\texttt{hrs\_cumulative\_multiplier\_9} & - & 1019 & 1150 \\
\texttt{hrs\_cumulative\_multiplier\_13} & - & 2322 & 2641 \\
\texttt{hrs\_cumulative\_multiplier\_17} & - & 4122 & 4625 \\
\texttt{modular\_adder\_4} & 17 & 17 & 23 \\
\texttt{modular\_adder\_6} & 36 & 37 & 46 \\
\texttt{modular\_adder\_8} & - & 69 & 80 \\
\texttt{modular\_adder\_10} & - & 111 & 131 \\
\texttt{modular\_adder\_12} & - & 165 & 231 \\
\texttt{modular\_adder\_14} & - & 223 & 290 \\
\texttt{modular\_adder\_16} & - & 303 & 398 \\
\texttt{modular\_adder\_18} & - & 395 & 521 \\
\texttt{modular\_adder\_20} & - & 493 & 668 \\
\texttt{multiplier\_4} & 30 & 30 & 32 \\
\texttt{multiplier\_12} & - & 661 & 925 \\
\texttt{multiplier\_20} & - & 2879 & 4675 \\
\texttt{qaoa\_3} & 8 & 8 & 8 \\
\texttt{qaoa\_5} & 34 & 35 & 37 \\
\texttt{qaoa\_6} & 40 & 40 & 47 \\
\texttt{qaoa\_7} & - & 75 & 96 \\
\texttt{qaoa\_8} & - & 94 & 111 \\
\texttt{qaoa\_9} & - & 121 & 168 \\
\texttt{qaoa\_10} & - & 189 & 236 \\
\texttt{qaoa\_11} & - & 218 & 267 \\
\texttt{qaoa\_12} & - & 271 & 360 \\
\texttt{qaoa\_13} & - & 250 & 309 \\
\texttt{qaoa\_14} & - & 420 & 520 \\
\texttt{qaoa\_15} & - & 398 & 552 \\
\texttt{qaoa\_16} & - & 476 & 665 \\
\texttt{qaoa\_17} & - & 611 & 779 \\
\texttt{qaoa\_18} & - & 619 & 875 \\
\texttt{qaoa\_19} & - & 762 & 1033 \\
\texttt{qaoa\_20} & - & 835 & 1167 \\
\texttt{qft\_3} & 11 & 11 & 14 \\
\texttt{qft\_4} & 19 & 19 & 23 \\
\texttt{qft\_5} & 30 & 30 & 39 \\
\texttt{qft\_6} & 45 & 45 & 53 \\
\texttt{qft\_7} & - & 66 & 75 \\
\texttt{qft\_8} & - & 89 & 98 \\
\texttt{qft\_9} & - & 112 & 128 \\
\texttt{qft\_10} & - & 141 & 163 \\
\texttt{qft\_11} & - & 171 & 205 \\
\texttt{qft\_12} & - & 206 & 261 \\
\texttt{qft\_13} & - & 254 & 294 \\
\texttt{qft\_14} & - & 287 & 377 \\
\texttt{qft\_15} & - & 347 & 401 \\
\texttt{qft\_16} & - & 401 & 489 \\
\texttt{qft\_17} & - & 460 & 514 \\
\texttt{qft\_18} & - & 495 & 651 \\
\texttt{qft\_19} & - & 553 & 650 \\
\texttt{qft\_20} & - & 609 & 715 \\
\texttt{qftentangled\_3} & 17 & 17 & 20 \\
\texttt{qftentangled\_4} & 27 & 27 & 32 \\
\texttt{qftentangled\_5} & 41 & 41 & 54 \\
\texttt{qftentangled\_6} & 58 & 58 & 75 \\
\texttt{qftentangled\_7} & - & 80 & 103 \\
\texttt{qftentangled\_8} & - & 103 & 144 \\
\texttt{qftentangled\_9} & - & 129 & 178 \\
\texttt{qftentangled\_10} & - & 161 & 243 \\
\texttt{qftentangled\_11} & - & 194 & 286 \\
\texttt{qftentangled\_12} & - & 234 & 334 \\
\texttt{qftentangled\_13} & - & 279 & 349 \\
\texttt{qftentangled\_14} & - & 328 & 457 \\
\texttt{qftentangled\_15} & - & 377 & 554 \\
\texttt{qftentangled\_16} & - & 431 & 524 \\
\texttt{qftentangled\_17} & - & 492 & 735 \\
\texttt{qftentangled\_18} & - & 546 & 680 \\
\texttt{qftentangled\_19} & - & 622 & 770 \\
\texttt{qftentangled\_20} & - & 698 & 882 \\
\texttt{qftentangled\_30} & - & 1448 & 1855 \\
\texttt{qnn\_3} & 4 & 4 & 4 \\
\texttt{qnn\_4} & 8 & 8 & 8 \\
\texttt{qnn\_5} & 10 & 10 & 12 \\
\texttt{qnn\_6} & 14 & 14 & 16 \\
\texttt{qnn\_7} & 16 & 16 & 20 \\
\texttt{qnn\_8} & 20 & 20 & 24 \\
\texttt{qnn\_9} & 22 & 22 & 28 \\
\texttt{qnn\_10} & 26 & 26 & 32 \\
\texttt{qnn\_11} & 28 & 28 & 36 \\
\texttt{qnn\_12} & 32 & 32 & 40 \\
\texttt{qnn\_13} & 34 & 34 & 44 \\
\texttt{qnn\_14} & 38 & 38 & 48 \\
\texttt{qnn\_15} & 40 & 40 & 52 \\
\texttt{qnn\_16} & 44 & 44 & 56 \\
\texttt{qnn\_17} & 46 & 46 & 60 \\
\texttt{qnn\_18} & 50 & 50 & 64 \\
\texttt{qnn\_19} & 52 & 52 & 68 \\
\texttt{qnn\_20} & 56 & 56 & 72 \\
\texttt{qpeexact\_3} & 4 & 4 & 4 \\
\texttt{qpeexact\_4} & 14 & 14 & 14 \\
\texttt{qpeexact\_5} & 29 & 29 & 33 \\
\texttt{qpeexact\_6} & 41 & 41 & 56 \\
\texttt{qpeexact\_7} & 56 & 56 & 68 \\
\texttt{qpeexact\_8} & - & 82 & 100 \\
\texttt{qpeexact\_9} & - & 105 & 122 \\
\texttt{qpeexact\_10} & - & 131 & 150 \\
\texttt{qpeexact\_11} & - & 168 & 182 \\
\texttt{qpeexact\_12} & - & 195 & 231 \\
\texttt{qpeexact\_13} & - & 230 & 277 \\
\texttt{qpeexact\_14} & - & 283 & 314 \\
\texttt{qpeexact\_15} & - & 319 & 403 \\
\texttt{qpeexact\_16} & - & 359 & 480 \\
\texttt{qpeexact\_17} & - & 428 & 490 \\
\texttt{qpeexact\_18} & - & 463 & 600 \\
\texttt{qpeexact\_19} & - & 515 & 590 \\
\texttt{qpeexact\_20} & - & 577 & 700 \\
\texttt{qpeinexact\_3} & 8 & 8 & 8 \\
\texttt{qpeinexact\_4} & 18 & 18 & 20 \\
\texttt{qpeinexact\_5} & 29 & 29 & 33 \\
\texttt{qpeinexact\_6} & 42 & 42 & 47 \\
\texttt{qpeinexact\_7} & 59 & 59 & 71 \\
\texttt{qpeinexact\_8} & - & 82 & 100 \\
\texttt{qpeinexact\_9} & - & 106 & 116 \\
\texttt{qpeinexact\_10} & - & 132 & 162 \\
\texttt{qpeinexact\_11} & - & 164 & 182 \\
\texttt{qpeinexact\_12} & - & 196 & 251 \\
\texttt{qpeinexact\_13} & - & 233 & 273 \\
\texttt{qpeinexact\_14} & - & 272 & 361 \\
\texttt{qpeinexact\_15} & - & 319 & 403 \\
\texttt{qpeinexact\_16} & - & 365 & 480 \\
\texttt{qpeinexact\_17} & - & 433 & 490 \\
\texttt{qpeinexact\_18} & - & 464 & 609 \\
\texttt{qpeinexact\_19} & - & 514 & 590 \\
\texttt{qpeinexact\_20} & - & 585 & 700 \\
\texttt{qwalk\_3} & 24 & 24 & 24 \\
\texttt{qwalk\_4} & - & 272 & 296 \\
\texttt{qwalk\_5} & - & 650 & 721 \\
\texttt{qwalk\_6} & - & 1903 & 2227 \\
\texttt{qwalk\_7} & - & 3377 & 3957 \\
\texttt{randomcircuit\_3} & 42 & 42 & 44 \\
\texttt{randomcircuit\_4} & 78 & 78 & 86 \\
\texttt{randomcircuit\_5} & - & 115 & 129 \\
\texttt{randomcircuit\_6} & - & 208 & 246 \\
\texttt{randomcircuit\_7} & - & 337 & 412 \\
\texttt{randomcircuit\_8} & - & 474 & 567 \\
\texttt{randomcircuit\_9} & - & 599 & 739 \\
\texttt{randomcircuit\_10} & - & 804 & 1028 \\
\texttt{randomcircuit\_11} & - & 955 & 1203 \\
\texttt{randomcircuit\_12} & - & 1183 & 1549 \\
\texttt{randomcircuit\_13} & - & 1487 & 1871 \\
\texttt{randomcircuit\_14} & - & 1841 & 2340 \\
\texttt{randomcircuit\_15} & - & 2144 & 2739 \\
\texttt{randomcircuit\_16} & - & 2613 & 3436 \\
\texttt{randomcircuit\_17} & - & 2689 & 3540 \\
\texttt{randomcircuit\_18} & - & 3233 & 4306 \\
\texttt{randomcircuit\_19} & - & 3885 & 5125 \\
\texttt{randomcircuit\_20} & - & 4355 & 5681 \\
\texttt{rg\_qft\_multiplier\_4} & 30 & 30 & 32 \\
\texttt{rg\_qft\_multiplier\_8} & - & 216 & 278 \\
\texttt{rg\_qft\_multiplier\_12} & - & 662 & 935 \\
\texttt{rg\_qft\_multiplier\_16} & - & 1497 & 2055 \\
\texttt{rg\_qft\_multiplier\_20} & - & 2914 & 4675 \\
\texttt{vbe\_ripple\_carry\_adder\_4} & 28 & 28 & 33 \\
\texttt{vbe\_ripple\_carry\_adder\_7} & - & 93 & 121 \\
\texttt{vbe\_ripple\_carry\_adder\_10} & - & 158 & 199 \\
\texttt{vbe\_ripple\_carry\_adder\_13} & - & 219 & 287 \\
\texttt{vbe\_ripple\_carry\_adder\_16} & - & 284 & 393 \\
\texttt{vbe\_ripple\_carry\_adder\_19} & - & 361 & 448 \\
\texttt{vqe\_real\_amp\_3} & 12 & 12 & 12 \\
\texttt{vqe\_real\_amp\_4} & 24 & 24 & 28 \\
\texttt{vqe\_real\_amp\_5} & 32 & 32 & 37 \\
\texttt{vqe\_real\_amp\_6} & 42 & 43 & 49 \\
\texttt{vqe\_real\_amp\_7} & 50 & 52 & 66 \\
\texttt{vqe\_real\_amp\_8} & 60 & 63 & 68 \\
\texttt{vqe\_real\_amp\_9} & 69 & 72 & 81 \\
\texttt{vqe\_real\_amp\_10} & - & 85 & 91 \\
\texttt{vqe\_real\_amp\_11} & - & 94 & 113 \\
\texttt{vqe\_real\_amp\_12} & - & 106 & 136 \\
\texttt{vqe\_real\_amp\_13} & - & 114 & 134 \\
\texttt{vqe\_real\_amp\_14} & - & 126 & 152 \\
\texttt{vqe\_real\_amp\_15} & - & 134 & 170 \\
\texttt{vqe\_real\_amp\_16} & - & 145 & 174 \\
\texttt{vqe\_real\_amp\_17} & - & 154 & 188 \\
\texttt{vqe\_real\_amp\_18} & - & 166 & 199 \\
\texttt{vqe\_real\_amp\_19} & - & 174 & 214 \\
\texttt{vqe\_real\_amp\_20} & - & 186 & 246 \\
\texttt{vqe\_su2\_3} & 12 & 12 & 12 \\
\texttt{vqe\_su2\_4} & 24 & 24 & 28 \\
\texttt{vqe\_su2\_5} & 32 & 32 & 37 \\
\texttt{vqe\_su2\_6} & 42 & 43 & 49 \\
\texttt{vqe\_su2\_7} & 50 & 52 & 66 \\
\texttt{vqe\_su2\_8} & - & 63 & 68 \\
\texttt{vqe\_su2\_9} & 69 & 72 & 81 \\
\texttt{vqe\_su2\_10} & - & 86 & 91 \\
\texttt{vqe\_su2\_11} & - & 94 & 113 \\
\texttt{vqe\_su2\_12} & - & 105 & 136 \\
\texttt{vqe\_su2\_13} & - & 114 & 134 \\
\texttt{vqe\_su2\_14} & - & 126 & 152 \\
\texttt{vqe\_su2\_15} & - & 134 & 170 \\
\texttt{vqe\_su2\_16} & - & 146 & 173 \\
\texttt{vqe\_su2\_17} & - & 154 & 188 \\
\texttt{vqe\_su2\_18} & - & 166 & 200 \\
\texttt{vqe\_su2\_19} & - & 174 & 213 \\
\texttt{vqe\_su2\_20} & - & 186 & 246 \\
\texttt{vqe\_two\_local\_3} & 28 & 28 & 36 \\
\texttt{vqe\_two\_local\_4} & 55 & 55 & 57 \\
\texttt{vqe\_two\_local\_5} & - & 90 & 101 \\
\texttt{vqe\_two\_local\_6} & - & 135 & 162 \\
\texttt{vqe\_two\_local\_7} & - & 192 & 238 \\
\texttt{vqe\_two\_local\_8} & - & 256 & 314 \\
\texttt{vqe\_two\_local\_9} & - & 335 & 386 \\
\texttt{vqe\_two\_local\_10} & - & 427 & 462 \\
\texttt{vqe\_two\_local\_11} & - & 523 & 669 \\
\texttt{vqe\_two\_local\_12} & - & 639 & 775 \\
\texttt{vqe\_two\_local\_13} & - & 756 & 954 \\
\texttt{vqe\_two\_local\_14} & - & 900 & 982 \\
\texttt{vqe\_two\_local\_15} & - & 1046 & 1387 \\
\texttt{vqe\_two\_local\_16} & - & 1190 & 1399 \\
\texttt{vqe\_two\_local\_17} & - & 1313 & 1628 \\
\texttt{vqe\_two\_local\_18} & - & 1526 & 2145 \\
\texttt{vqe\_two\_local\_19} & - & 1688 & 2318 \\
\texttt{vqe\_two\_local\_20} & - & 1898 & 2622 \\
\texttt{wstate\_3} & 8 & 8 & 8 \\
\texttt{wstate\_4} & 14 & 14 & 18 \\
\texttt{wstate\_5} & 20 & 20 & 22 \\
\texttt{wstate\_6} & 26 & 26 & 28 \\
\texttt{wstate\_7} & 32 & 32 & 41 \\
\texttt{wstate\_8} & 38 & 38 & 42 \\
\texttt{wstate\_9} & 44 & 44 & 50 \\
\texttt{wstate\_10} & 50 & 50 & 65 \\
\texttt{wstate\_11} & 56 & 56 & 60 \\
\texttt{wstate\_12} & 62 & 62 & 70 \\
\texttt{wstate\_13} & 68 & 68 & 105 \\
\texttt{wstate\_14} & 74 & 74 & 122 \\
\texttt{wstate\_15} & - & 80 & 88 \\
\texttt{wstate\_16} & - & 86 & 94 \\
\texttt{wstate\_17} & - & 92 & 102 \\
\texttt{wstate\_18} & - & 98 & 108 \\
\texttt{wstate\_19} & - & 104 & 112 \\
\texttt{wstate\_20} & - & 110 & 122 \\
\end{longtblr}

}